\documentstyle[sprocl,psfig]{article}
\def\to{\rightarrow}
\def\bq{\begin{equation}}
\def\eq{\end{equation}}
\def\ba{\begin{eqnarray}}
\def\ea{\end{eqnarray}}
\def\lsim{\mathrel{\raisebox{-.6ex}{$\stackrel{\textstyle<}{\sim}$}}}
\def\gsim{\mathrel{\raisebox{-.6ex}{$\stackrel{\textstyle>}{\sim}$}}}

\def\epem{e^+e^-}
\def\wpwm{W^+W^-}

\def\tptm{\tau^+\tau^-}
\newcommand{\sla}[1]{/\!\!\!#1}%
\begin{document}

\vspace*{-.5in}

\font\fortssbx=cmssbx10 scaled \magstep1
\hbox to \hsize{
\includegraphics{/NextLibrary/TeX/tex/inputs/uwlogo.ps}
\hskip.25in \raise.05in
\hbox{\fortssbx University of Wisconsin - Madison}
\hfill\vbox{\hbox{\bf MADPH-99-1100}
            \hbox{February 1999}} }

\bigskip

\title{COLLIDER PHYSICS\footnote{Lectures presented at the Theoretical
Advanced Study Institute (TASI), Boulder, CO, June 1998. }
}

\author{\unskip\bigskip DIETER ZEPPENFELD}

\address{Department of Physics, University of Wisconsin,\\ 1150 University 
Avenue, Madison, WI 53706, USA\\ E-mail: dieter@pheno.physics.wisc.edu}


\maketitle\abstracts{
These lectures are intended as a pedagogical introduction to physics at 
$e^+e^-$ and hadron colliders. A selection of processes is used to 
illustrate the strengths and capabilities of the different machines. 
The discussion includes $W$ pair production and chargino searches 
at $e^+e^-$ colliders, Drell-Yan events and the top quark search at the 
Tevatron, and Higgs searches at the LHC.
}

\section{Introduction}
\label{sec:intro}

Over the past two decades particle physics has advanced to the stage where
most, if not all observed phenomena can be described, at least in principle,
in terms of a fairly economic $SU(3)\times SU(2)\times U(1)$ gauge theory:
the Standard Model (SM) with its three generations of quarks and 
leptons.\cite{altar}
Still, there are major short-comings of this model. The spontaneous breaking
of the electroweak sector is parameterized in a simple, yet ad-hoc manner,
with the help of a single scalar doublet field. The smallness of the 
electroweak scale, $v=246$~GeV, as compared to the Planck scale, requires
incredible fine-tuning of parameters. The Yukawa couplings of the fermions
to the Higgs doublet field, and thus fermion mass generation, cannot be 
further explained within the model. For all these reasons there is a strong
conviction among particle physicists that the SM is an effective theory
only, valid at the low energies presently being probed, but eventually
to be superseded by a more fundamental description.

\looseness=-1
Extensions of the SM include models with a larger Higgs 
sector,\cite{GHKD,Peskin} e.g. in the
form of extra doublets giving rise to more than the single scalar Higgs
resonance of the SM, models with extended gauge groups which predict e.g.
extra $Z$-bosons to exist at high energies,\cite{cvetic} 
or supersymmetric extensions of the SM with a doubling of all known 
particles to accommodate the extra bosons and fermions required by 
supersymmetry.\cite{mssm} In all these cases new heavy
quanta are predicted to exist, with masses in the 100~GeV region and beyond. 
One of the main goals of present particle physics is to discover unambiguous
evidence for these new particles and to thus learn experimentally what lies
beyond the~SM.

While some information on physics beyond the SM can be gleaned from precision 
experiments at low energies,\cite{altar} 
via virtual contributions of new heavy particles
to observables such as the anomalous magnetic moment of the muon, rare
decay modes of heavy fermions (such as $b\to s\gamma$), or the so-called
oblique corrections to electroweak observables, a complete 
understanding of what lies beyond the SM will require the direct production
and detailed study of the decays of new particles. 

The high center of mass energies required to produce these massive objects 
can only be generated by colliding beams of particles, which in practice means
electrons and protons\footnote{The feasibility of a $\mu^+\mu^-$ collider is
under intense study as well.\cite{mupmum} For the purpose of these lectures, 
the physics investigations at such machines would be very similar to the ones 
at $e^+e^-$colliders.}.
$e^+e^-$ colliders have been operating for almost thirty years, the highest
energy machine at present being the LEP storage ring at CERN with a center of 
mass energy close to 200~GeV. Proton-antiproton collisions have been the 
source of the highest energies since the early 1980's, with first the CERN 
collider and now the Tevatron at Fermilab, which will continue operation 
with a center of mass (c.m.) energy of 2~TeV in the year 2000. 
In 2005 the Large Hadron Collider (LHC) in the LEP tunnel 
at CERN will raise the maximum energy to 
14~TeV in proton-proton collisions. Somewhat later we may see a linear 
$e^+e^-$ collider with a center of mass energy close to 1~TeV. 

The purpose of these lectures is 
to describe the research which can be performed with these existing, or soon 
to exist machines at the high energy frontier. No attempt will be made to 
provide complete coverage of all the questions which can be addressed 
experimentally at these machines. Books have been written for this 
purpose.\cite{collphys} Rather I will use a few specific examples to
illustrate the strengths (and the weaknesses) of the various colliders,
and to describe a variety of the analysis tools which are being used in 
the search for new physics. Section~\ref{sec:overview} starts with an overview
of the different machines, of the general features of new particle production
processes, and of the ensuing implications for the detectors which are used
to study them. This section includes a theorist's picture of the structure
of modern collider detectors.

Specific processes at $e^+e^-$ and hadron colliders are discussed in the 
three main Sections of these lectures. Section~\ref{sec:ee} first deals 
with $W^+W^-$ production in $e^+e^-$ collisions, and how this process is 
used to measure the $W$-mass. It then discusses chargino pair production 
as an example for new particle searches in the relatively clean environment 
of $e^+e^-$ colliders.
Hadron collider experiments are discussed in Sections~\ref{sec:pp} and
\ref{sec:Higgs.at.LHC}. After consideration of the basic structure of cross 
sections at hadron colliders and the need for non-perturbative input in 
the form of parton distribution functions, the simplest new physics process, 
single gauge boson (Drell-Yan) production will be considered in some detail. 
The top quark search at the Tevatron is used as an example for a successful 
search for heavy quanta. Section~\ref{sec:Higgs.at.LHC} then looks at
LHC techniques for finding the Higgs boson.
Some final conclusions are drawn in Section~\ref{sec:conclusions}.

\section{Overview}
\label{sec:overview}

Some features of production processes for new heavy particles are fairly
general. They are important for the design of colliding
beam accelerators and detectors, because they concern the angular and energy 
dependence of pair-production processes. 

In order to extend the reach for producing new heavy particles, the available
center of mass energy, at the quark, lepton or gluon level, is continuously
being increased, in the hope of crossing production thresholds. In turn this 
implies that new particles will be discovered close to pair-production 
threshold, where their momenta are still fairly small compared to their mass. 
Small momenta, however, imply little angular dependence of matrix elements
because all $\cos\theta$ or $\sin\theta$ dependence is suppressed by powers 
of $\beta=|{\bf p}|/p^0$. As a result the production of heavy particles is 
fairly isotropic in the center of mass system. This is simple quantum 
mechanics, of course; higher multipoles $L$ in the angular distribution
are suppressed by factors of $\beta^{2L+1}$.

\begin{figure}[t]
\centering\leavevmode
\psfig{figure=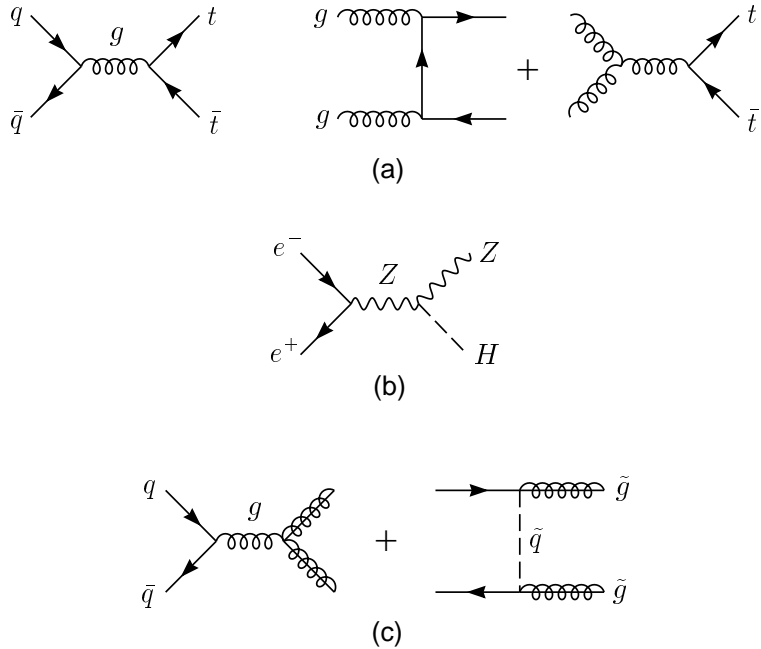,height=3.5in}
\vspace*{-0.1in}
\caption{
Representative Feynman graphs for new particle production processes, 
(a) top quark production, $q\bar q \to t\bar t$ and $gg\to t\bar t$,
(b) Higgs production, $e^+e^-\to ZH$, and
(c) gluino pair production, $q\bar q \to \tilde g\tilde g$.
\label{fig:feynman}
}
\vspace*{-0.05in}
\end{figure}

Another feature follows from the fact that pair production processes have a 
dimensionless amplitude, which can only depend on ratios like 
$M^2/\hat s$, where $M$ is the mass of the produced particles and 
$\sqrt{\hat s}$ is the available center of mass energy. 
At fixed scattering angle, the amplitude approaches a constant as
$M^2/\hat s\to 0$. The effect is most pronounced when
the pair production process is dominated by gauge-boson exchange in the 
$s$-channel as is the case for a large class of new particle searches, be it
$q\bar q \to t\bar t$ at the Tevatron, $e^+e^-\to ZH$ 
at LEP2, or $q\bar q$ annihilation to two gluinos at the LHC (see 
Fig.~\ref{fig:feynman}). $s$-channel production allows $J=1$ final states 
only, which limits the angular dependence of the production cross section
to a low order polynomial in $\cos\theta$ and $\sin\theta$. We thus have 
little variation of the production amplitude, i.e.
\bq
{\cal M}(\hat s)\approx {\rm constant}\;,
\eq
where the constant is determined by the coupling constants at the vertices 
of the Feynman graphs of Fig.~\ref{fig:feynman}.

\begin{figure}[t]
\centering\leavevmode
\psfig{figure=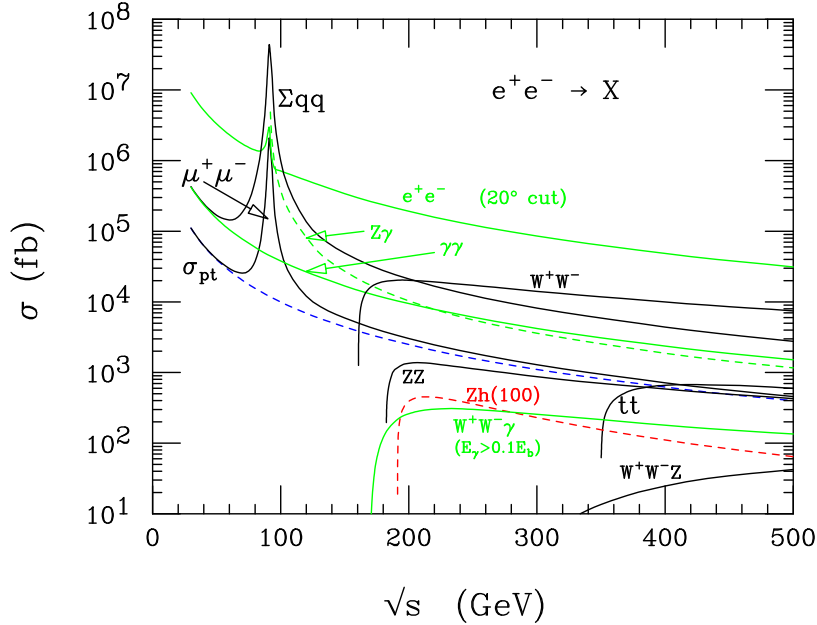,height=3.3in,angle=90}
\vspace*{-0.1in}
\caption{Representative SM production cross sections in $e^+e^-$ annihilation.
For Bhabha scattering and final states involving photons, the differential 
cross sections $d\sigma/d\cos\theta$ have been integrated with a cut of 
$20^o<\theta<160^o$, in order to avoid the singular forward region.
Note the $Z$ peak and the characteristic $1/s$ fall-off of cross
sections at high energies. 
\label{fig:sigmaSMee}
}
\vspace*{-0.05in}
\end{figure}

Since the production cross section is given by
\bq
{d\hat\sigma\over d\Omega} = {1\over 2\hat s}{\beta\over 32\pi^2}
\overline{\sum_{\rm pol}} |{\cal M}|^2 \;,
\eq
approximately constant ${\cal M}$ implies that 
the production cross section drops like $1/\hat s$ and becomes small fast 
at high energy, being of order $\alpha^2/\hat s\sim 1$~pb at 
$\sqrt{\hat s}=200$~GeV for electroweak strength cross sections. This $1/s$
fall-off of production cross sections is clearly visible in 
Fig.~\ref{fig:sigmaSMee}, above the peak at 91.2~GeV, produced by 
the $Z$ resonance. The rapid decrease of cross sections with energy 
implies that in order to search for new heavy particles one needs both
high energy and high luminosity colliders, with usable center of mass 
energies of order several hundred GeV and luminosities of order 
1~fb$^{-1}$ per year or higher.

The most direct and cleanest way to provide these high energies and 
luminosities is in $e^+e^-$ collisions. For most of the nineties, LEP at
CERN and the SLAC linear collider (SLC) have operated on the $Z$ peak, at
$\sqrt{s}\approx m_Z=91.187$~GeV, collecting some $10^7$ $Z$ decays at LEP
and several hundred thousand at SLC. 
The high counting rate provided by the $Z$ resonance (see 
Fig.~\ref{fig:sigmaSMee}) has allowed very precise measurements of the 
couplings of the $Z$ to the SM quarks and leptons. At the same time 
searches for new particles were conducted, and the fact that nothing was 
found excludes any new particle which could be pair-produced in $Z$ decays, 
i.e. which have normal gauge couplings to the $Z$ and have masses 
$M\lsim m_Z/2$. This includes fourth generation quarks and leptons and the
charginos, sleptons and squarks of supersymmetry.\cite{altar}

Since 1996 the energy of LEP has been increased in steps, via 161, 172,  and 
183~GeV, to 189~GeV in 1998. About 250~pb$^{-1}$ of data have been collected 
by each of the four LEP experiments over this period, mostly at the highest
energy point. The experiments have mapped out the $W^+W^-$ production 
threshold (see Fig.~\ref{fig:sigmaSMee}), measured the $W$ mass and $W$ 
couplings directly, searched for the Higgs and set new mass limits on other 
new particles (see Section~\ref{sec:ee}). 

Why has progress been so incremental? The main culprit is synchrotron 
radiation in circular machines. The centripetal acceleration of the
electrons on their circular orbit leads to an energy loss which grows as
$\gamma^4/R$, where $R$ is the bending radius of the machine, approximately
4.2~km at LEP, and $\gamma=E_e/m_e$ is the time dilatation factor
for the electrons, which at LEP exceeds $10^5$. The small
electron mass requires very large bending radii, which can be achieved with
very modest bending magnets in the synchrotron. Scaling up LEP to higher 
energies soon leads to impossible numbers. A 1~TeV electron synchrotron with 
the same synchrotron radiation loss as LEP (about 2.8~GeV per turn at 
$\sqrt{s}=200$~GeV)\cite{keil} would have to be over 600 times larger. 
The future of $e^+e^-$ machines belongs to linear colliders, which do not
reuse the accelerated electrons and positrons, but rather collide the beams 
once, in a very high intensity beam spot. The SLC has been the first 
successful
machine of this type. The next linear collider (NLC) would be a 400~GeV
to 1~TeV $e^+e^-$ collider with a luminosity in the 10--100~fb$^{-1}$ per
year range. With continued cooperation of physicists worldwide, such a 
machine might become a reality within a decade.

The easiest way to get to larger center of mass energies is to use
heavier beam particles, namely protons or anti-protons. Their 2000 times
larger mass makes synchrotron radiation losses negligible, even for much 
higher beam energies. Thus, the energy of proton storage rings is limited 
by the maximum magnetic fields which can be achieved to keep the particles 
on their circular orbits, i.e. the beam momentum $p$ is limited by the relation
\bq
p = eBR\; .
\eq
The Tevatron at Fermilab is the highest energy $\bar pp$ collider at present, 
and so far has accumulated about 120~pb$^{-1}$ of data, at a $\bar pp$ 
center of mass energy of 1.8~TeV, in each of two experiments, 
CDF and D0. It was this data taking period in the mid-nineties, called
run I, which led to the discovery of the top-quark (see Section~\ref{sec:pp}).
Note that the pair production of top-quarks, with a mass of $m_t=175$~GeV, 
will be possible at $e^+e^-$ colliders only in the NLC era. A higher 
luminosity run at 2~TeV, run II, is scheduled to collect 1--2~fb$^{-1}$ of 
data, starting in the spring of 2000. And before an NLC will be constructed, 
the LHC at CERN will start with $pp$ collisions at a c.m. energy of 14~TeV 
and with a luminosity of $10^{33}$--$10^{34}\;{\rm cm}^{-2}{\rm sec}^{-1}$, 
corresponding to 10--100~fb$^{-1}$ per year.

Given these much higher energies available at hadron colliders, why do we
still invest in $e^+e^-$ machines? The problem with hadron colliders is that
protons are composite objects, made out of quarks and gluons, and these partons
only carry a fraction of the proton energy. In order to produce new heavy 
particles the c.m. energy in a parton-parton collision must be larger than
the sum of the masses of the produced particles, and this becomes increasingly
unlikely as the required energy exceeds some 10--20\% of the collider energy.
Most $pp$ or $\bar pp$ collisions are collisions between fairly low energy 
partons. Since the proton is a composite object, of finite size, the total
$pp$ or $\bar pp$ cross section does not decrease with energy, in fact it 
grows logarithmically and reaches about 100~mb at LHC energies. The high 
energy parton-parton collisions, however, suffer from the $1/s$ suppression
discussed before. Production cross sections for new particles are of order
1~pb (with large variations), i.e. $10^{-11}$ times smaller. At hadron 
colliders, one thus needs to identify one interesting event in a background of
$10^{11}$ bad ones, which poses a daunting task to the experimentalists and 
their detectors. The much cleaner situation at $e^+e^-$ colliders can be 
appreciated from Fig.~\ref{fig:sigmaSMee}: backgrounds to new physics searches
arise form processes like $e^+e^-\to q\bar q$, $e^+e^-\to W^+W^-$, or
$e^+e^-\to Z\gamma$, with cross sections in the 1--100~pb region and thus
not much larger than the expected signal cross sections.

\begin{figure}[t]
\centering\leavevmode
\psfig{figure=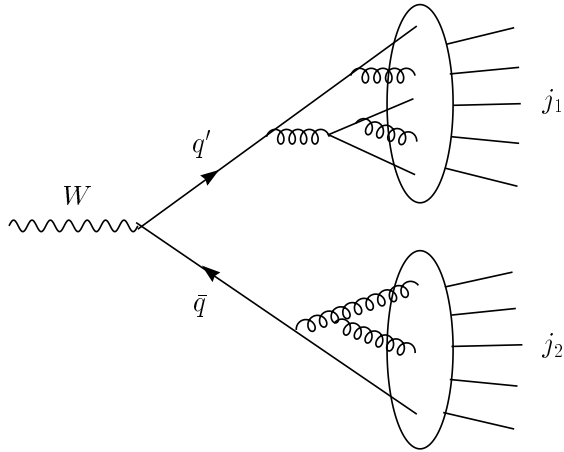,height=2.5in}
\vspace*{-0.1in}
\caption{Schematic drawing representing decay of a $W^+\to \bar q q$ with
subsequent parton showering and hadronization, leading to a dijet signature
for the $W$ in the detector. 
\label{fig:Wdecay}
}
\vspace*{-0.1in}
\end{figure}

Whether new particles are produced in $e^+e^-$ or $pp$ collisions, they are
never expected to be seen directly in the detector. Expected lifetimes, 
$\tau$, and ensuing decay lengths, $\gamma c\tau$, are of nuclear scale and, 
therefore, only the decay products can be observed. A typical example is 
shown in Fig.~\ref{fig:Wdecay}. A $W$ boson has a decay width of 
$\Gamma_W\approx 2.1$~GeV
and is depicted to decay into a $u\bar d$ quark pair (other decay modes
are $e^+\nu_e,\; \mu^+\nu_\mu,\; \tau^+\nu_\tau$ and $c\bar s$). The produced
quarks, of course, are not observable either, due to confinement, rather they
emit gluons and quark-antiquark pairs, which eventually hadronize and form jets
of hadrons containing pions, kaons, and so on. The detectors have to observe 
these hadrons, measure the directions and energies of the jets, and deduce 
from here the four-momenta of the original $W^+$ or $u$ and $\bar d$ quarks.

The situation is very similar for other new heavy quanta. A Higgs boson, of
mass $m_H=120$~GeV say, is expected to decay into $b\bar b$, $\tau^+\tau^-$
or $\gamma\gamma$, among others, with expected branching ratios
\ba
B(H\to b\bar b) & = & 74\% \\
B(H\to \tau^+\tau^-) & = & 6\% \\
B(H\to \gamma\gamma) & = & 0.2\% \;.
\ea
At LEP2, $ZH$ production is searched for in the $b\bar b$ decay mode of the 
Higgs, i.e. $b$-quark jets need to be observed and distinguished from
lighter quark jets. 

Supersymmetric particles are expected to produce an entire decay chain before
they can be observed in the detector, an example being the
decay of a gluino to a squark and a quark, where the squark in turn decays
to the lightest neutralino, $\tilde\chi^0$ or chargino, $\tilde\chi^\pm$,
\ba
\tilde g &\to& \tilde q\bar q \to \bar q q \tilde\chi^0\to jj\sla E_T \;,\\
\tilde g &\to& \tilde q\bar q \to \bar q q' \tilde\chi^\pm
	\to \bar q q' W^\pm \tilde\chi^0 \to \bar q q'\ell^\pm\nu\tilde\chi^0
        \to jj \ell^\pm\sla E_T \; .
\ea
In the last step of the decay chain, hadronization of the quarks leads to 
jets, and the neutrino and the neutralino escape the detector, leading 
to an imbalance in the measured momenta of observable particles transverse 
to the beam, i.e. to a missing $E_T$ signature.

\begin{figure}[t]
\centering\leavevmode
\psfig{figure=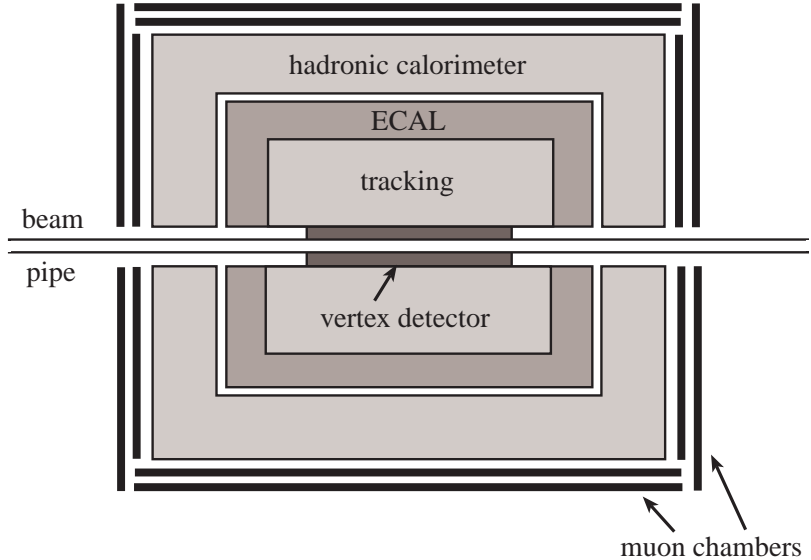,height=3.0in}
\vspace*{-0.1in}
\caption{Schematic cross section through a typical collider detector. Shown, 
from the beam pipe out, are the locations of vertex detector, tracker,
electromagnetic calorimeter, hadronic calorimeter, and muon chambers.
\label{fig:toy.detector}
}
\vspace*{-0.1in}
\end{figure}

Collider detectors, which are to discover these new particles, must be designed
to observe the decay products, positively identify them and measure their
direction and their energy in the lab, i.e. determine 
the momenta of electrons, muons,
photons and jets. In addition, the identification of heavy quarks, in 
particular the $b$ quarks arising in Higgs boson and top quark decays, 
is very important. Many technical solutions have been developed for this 
purpose, variations in the type of detector doing the reconstruction of 
charged particle tracks, or the energy measurement of electrons and photons 
(via electromagnetic showers) or jets (in a hadronic calorimeter). The basic
layout of modern collider detectors is remarkably uniform, however, for 
fundamental physics reasons. For the basic level addressed in these 
lectures, it is sufficient to have a brief look at these common, global
features. 

The schematic drawing of Fig.~\ref{fig:toy.detector} shows a cross section 
through a typical collider detector. The detector has a cylindrical 
structure, which is wrapped around the beam pipe in which the electrons or
protons collide, at the center of the detector. The standard components of 
the detector are as follows.
 
\begin{itemize}
 
\item{} Central Tracking: The innermost part of the
detector, closest to the interaction region, records the tracks of the produced
charged particles, $e^\pm$, $\mu^\pm$, $\pi^\pm$, $K^\pm$ and so on. 
Via the bending of the tracks in a strong magnetic field, pointing along 
the beam direction, it measures the charged particle momenta. 

\item{} Electromagnetic Calorimeter: Neutral 
particles leave no tracks. They need to be measured via total absorption of 
their energy in a calorimeter. Electrons and photons loose their energy 
relatively quickly, via an electromagnetic shower. Since a shower develops 
randomly, statistical fluctuations limit the accuracy of the energy 
measurement. Excellent results, such as for the CMS detector at the 
LHC,\cite{CMS} are
\bq
{\Delta E  \over E} = {0.02\over\sqrt{E}}\quad {\rm to} \quad
{0.05\over\sqrt{E}}\; ,
\eq
where the energy $E$ is measured in GeV. Note that neutral pions decay into 
two photons before they leave the beam pipe, and these photons will stay
very close to each other for large $\pi^0$ momentum. Hence, $\pi^0$s will 
be recorded in the electromagnetic calorimeter and they can fake photons.

\item{} Hadronic Calorimeter: Other hadrons are absorbed in the hadron 
calorimeter where their energy deposition is measured, with a statistical 
error which may reach~\cite{ATLAS}
\bq
{\Delta E  \over E} = {0.4\over \sqrt{E} }\; .
\eq
Typical hadron calorimeters have a thickness of some 25 absorption lengths
and normally only muons, which do not interact strongly with the heavy nuclei
of the hadron calorimeter, will penetrate it. Outside the calorimeter one 
therefore places the 

\item{} Muon Chambers: They record the location where a penetrating particle
leaves the inner detector, and in several layers follows the direction of the 
track. Together with the central tracking information, this allows to 
measure the curvature of the muon track in the known 
magnetic fields inside the detector and determines the muon momentum.

\item{} Vertex Detector: Bottom and charm quarks can be identified by the 
finite lifetime, of order 1~ps, of the hadrons which they form. This lifetime
leads to a decay length of up to a few mm, i.e. the $b$ or $c$ decay products
do not point back to the primary interaction vertex, which is much smaller, 
but to a secondary vertex. The decay length can be resolved with very high
precision tracking. For this purpose modern collider detectors possess a 
solid state micro-vertex detector, very close to the beam-pipe, which provides
information on the location of tracks with a resolution of order 10~$\mu$m.
This technique now allows to identify centrally produced $b$-quarks, i.e. 
those that are produced at angles of more than a few degrees with respect 
to the beams, with efficiencies above 50\% and with high purity, rejecting 
non-$b$ jets with more than 95\% probability.
 
\end{itemize}

The various elements of the detector work together to identify the components
of an event. An electron would deposit its energy in the electromagnetic 
calorimeter and produce a central track, which distinguishes it from a neutral
photon. A charged pion or kaon produces a track also, but only a fraction 
of its energy ends up in the electromagnetic calorimeter: most of it leaks 
into the hadronic calorimeter. A muon, finally, deposits little energy
in either calorimeter, rather it leads to a central track and to hits in the 
muon chambers. 

If this muon originates from a $b$-decay, it will be traveling
in the same direction as other hadrons belonging to the $b$-quark jet. This
muon is not isolated, as opposed to a muon from a decay $W\to\mu\nu$ which
only has a small probability to travel into the same direction as the twenty
or thirty hadrons of a typical jet. One thus obtains a very detailed picture
of the entire event, and from this picture one needs to reconstruct what 
happened at the parton level.

\section{$e^+e^-$ Colliders}
\label{sec:ee}

A full event reconstruction is most easily done at an $e^+e^-$ 
collider where the beam particles are elementary objects, i.e. the 
entire energy of the collision can go into the production of heavy 
particles. In contrast,
at hadron colliders, the additional partons in the parent protons lead to a
spray of hadrons in the detector which obscure the parton-parton collision
we are interested in. 

In $e^+e^-$ collisions at energies $\sqrt{s}\lsim 160$~GeV, the dominant hard 
process is fermion pair production, $e^+e^-\to \bar ff$. For $f=e,\;\mu$ this 
leads to extremely clean events with two particles in the final state only,
and even the bulk of $\bar q q$ production events are easily recognized as 
dijet events. With the advent of LEP2 the situation has become more 
complicated. As can be seen in Fig.~\ref{fig:sigmaSMee}, $W^+W^-$ and $ZZ$ 
events are as copious as fermion pair production, but they lead to a more 
complex 4-fermion final state, via the decay of the two $W$s or $Z$s into a 
pair of quarks or leptons each. Let us start our survey of $e^+e^-$ physics
with this new class of events, which will be important in all new high energy
$\epem$ colliders. They show interesting features and provide information 
on fundamental parameters of the SM in their own right, but also they form 
an important background to new particle searches, such as charginos for 
example, and we need to analyze them in some detail. 

\begin{figure}[b]
\centering\leavevmode
\psfig{figure=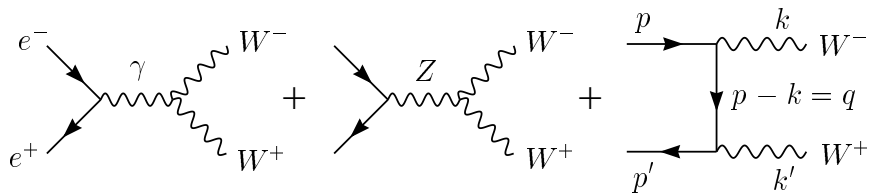,height=1.1in}
\vspace*{-0.1in}
\caption{
Feynman graphs for the process $e^+e^-\to W^+W^-$. The labels on the neutrino
exchange graph give the particle momenta.
\label{fig:feynman.ww}
}
\vspace*{-0.1in}
\end{figure}

\subsection{$W^+W^-$ production} 
\label{sec:ee.ww}

At tree level, three Feynman graphs contribute to the production of $W$ 
pairs.\cite{hpzh,lep2ww}
They are $s$-channel $\gamma$- and $Z$-exchange, and $t$-channel neutrino
exchange, and are shown in Fig.~\ref{fig:feynman.ww}. The dominant features 
of the production cross section can be read off the propagator structure 
of the individual graphs. While the two $s$-channel amplitudes show a modest 
dependence on scattering angle only, the neutrino exchange graph produces 
a strong peaking at small scattering angles: the propagator factor for this
graph is $1/t$ with $t = (p-k)^2=-2p\cdot k+m_W^2$. 
In the c.m. system we choose
the initial electron momentum, $p$, along the $z$-axis and define the 
scattering angle as the angle between the $W^-$ and the $e^-$ direction, i.e.
\bq
p={\sqrt{s}\over 2}(1,0,0,1)\;, \qquad 
k={\sqrt{s}\over 2}(1,\beta\sin\theta,0,\beta\cos\theta)\; ,
\eq
where $\beta=\sqrt{1-4m_W^2/s}$ is the velocity of the produced $W$. The 
denominator of the propagator then becomes 
\bq
|t| = {s\over 4}\left(2-2\beta\cos\theta-{4m_W^2\over s}\right) = 
{s\over 4}\left(1+\beta^2-2\beta\cos\theta\right) \ge 
{s\over 4}\left(1-\beta\right)^2 \; .
\eq

\begin{figure}[t]
\centering\leavevmode
\psfig{figure=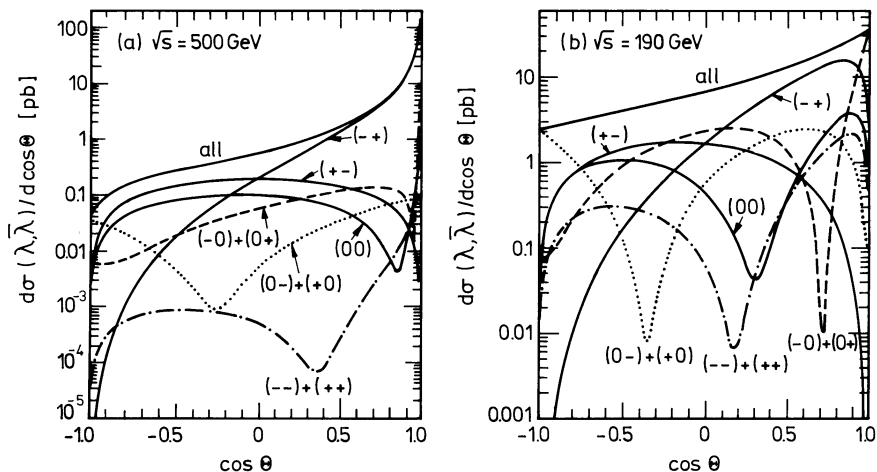,height=2.5in}
\vspace*{-0.1in}
\caption{
Angular distribution $d\sigma/d\cos\theta$ for $e^+e^-\to W^+W^-$ production 
at (a) $\protect\sqrt{s}=500$~GeV and (b) $\protect\sqrt{s}=190$~GeV, 
within the SM. Also
shown are the cross sections for the various $W^-W^+$ helicity combinations
$(\lambda,\bar\lambda)$. Where a sum is shown, both helicity combinations 
give the same result. From Ref.~\protect\cite{hpzh}.
\label{fig:dsdc.ww.pol}
}
\vspace*{-0.1in}
\end{figure}

Close to threshold, i.e. for $\beta\approx 0$, there is little angular 
dependence. At high energies, however, as $\beta\to 1$, $|t|$ becomes very 
small near $\theta=0$ and the $1/t$ pole induces a strongly peaked $W$ pair 
cross section near forward scattering angles. This effect is shown in 
Fig.~\ref{fig:dsdc.ww.pol}, where the angular distribution
$d\sigma/d\cos\theta$ for $W^+W^-$ production in the SM is shown, including
the contributions from different $W^-$ and $W^+$ helicities, $\lambda$ and
$\bar\lambda$. Angular momentum conservation
and the fact that left-handed electrons only contribute to the $\nu$-exchange
graph, lead to a strong polarization of the produced $W$'s in the forward 
region: the produced $W^-$ helicity is mostly $-1$ in this region, i.e. it
picks up the electron helicity.

The $W$s are only observed via their decay products, $W\to\bar ff'$, where
three $\ell\nu$ combinations $(\ell=e,\mu,\tau)$ and six quark combinations 
can be produced ($u\bar d$ and $c\bar s$ and counting $N=3$ different colors). 
Since all quarks and leptons couple equally to $W$s (when neglecting 
Cabibbo mixing), and because any CKM effects exactly
compensate in the decay widths, the branching ratios to leptons and hadrons 
are simply given by
\ba\label{eq:B.W.1}
B(W\to e\nu_e)=B(W\to\mu\nu_\mu)&=&B(W\to\tau\nu_\tau)={1\over 9}\;, \\
B(W\to {\rm hadrons}) &=& {1\over 9}\cdot 2\cdot 3 = {2\over 3}\; .
\label{eq:B.W.2}
\ea
QCD effects induce corrections of a few percent to these relations. 
In lowest order, a decay $W\to \bar qq'$ leads
to two jets in the final state, and this, combined with the branching ratios
of Eq.~(\ref{eq:B.W.1},\ref{eq:B.W.2}), fixes the probabilities for the 
various classes of events to be observed for $W^+W^-$ production,
\ba
B(W^+W^-\to {4 \rm jets}) &=& 46\%\;, \\
B(W^+W^-\to jj+e\nu_e,\mu\nu_\mu) &=& 29\%\;, \\
B(W^+W^-\to \ell^+\nu\ell^-\bar\nu) &=& 10.5\%\; , \\
B(W^+W^-\to jj+\tau\nu_\tau) &=& 14.5\%\; .
\ea
Thus, it is most likely to observe the two $W$s in 4-jet events, followed 
by the `semileptonic' channel, where one $W$ decays into either electrons
or muons. The remaining channels have at least two neutrinos in the final
state (the $\tau$ decays inside the beam-pipe!) and hence a substantial 
fraction of the final state particles cannot be observed, which limits 
the reconstruction of the event. Fortunately, these more difficult situations
comprise only one quarter of the $W$ pair sample.

\begin{figure}
\centering\leavevmode
\psfig{figure=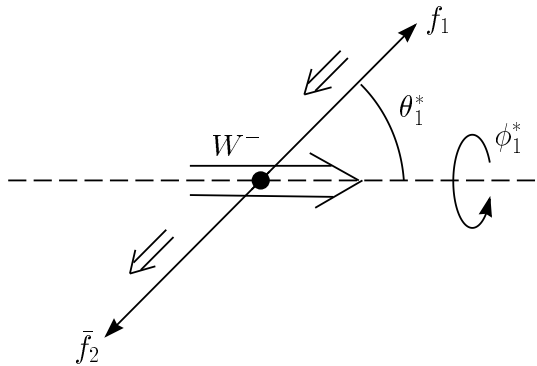,height=2.0in}
\caption{
Orientation of spins for the decay of a right-handed $W^-$ into a pair
of fermions, $f_1\bar f_2$. Because of the $V-A$ structure of charged
currents, the fermion helicities are fixed as shown.
\label{fig:Wpol}
}
\vspace*{-0.1in}
\end{figure}

As we saw when discussing production angular distributions, the $W$s are 
strongly polarized. Fortunately, the $V-A$ structure of the $W$-fermion 
couplings provides a very efficient polarization analyzer for the $W$s, via 
their decay distributions. Consider the decay of a right-handed $W^-$, i.e.
of helicity $\lambda=+1$, as depicted in Fig.~\ref{fig:Wpol} in the 
$W$ rest frame. Because of the $V-A$ coupling, the decay $W\to f_1\bar f_2$
always leads to a left-handed fermion and a right-handed anti-fermion, in the
massless fermion limit. The fermion spins therefore always line up as shown
in the figure. Taking the decay polar angle $\theta_1^*\to 0$, the combined 
fermion spins point opposite to the spin of the parent $W^-$. Angular momentum
conservation does not allow this, which means that the decay amplitude 
vanishes at $\cos\theta_1^*= 1$. The same
argument shows that for a left-handed $W^-$, of helicity $\lambda=-1$, the
decay amplitude must vanish for $\cos\theta_1^*=-1$. A quick calculation shows
that the decay amplitudes for $W^-\to f_1\bar f_2$ are proportional to
\ba
D(W^-\to f_1\bar f_2)(\lambda=+1) & = & 
{1\over \sqrt 2}\left(1-\cos\theta_1^*\right)e^{i\phi_1^*} \;, \\
D(W^-\to f_1\bar f_2)(\lambda=0) & = & -\sin\theta_1^* \;, \\
D(W^-\to f_1\bar f_2)(\lambda=-1) & = & 
{1\over \sqrt 2}\left(1+\cos\theta_1^*\right)e^{-i\phi_1^*} \;, 
\ea
where $\phi_1^*$ is the decay azimuthal angle, as indicated in the figure.
Analogous results are obtained, of course, for the $W^+$ decay angles
$\theta_2^*$ and $\phi_2^*$. 

By measuring the decay angular distributions, i.e. by distinguishing the
$(1-\cos\theta^*)^2$, $\sin^2\theta^*$, and  $(1+\cos\theta^*)^2$ distributions
for right-handed, longitudinal and left-handed polarized $W$s, we can measure
the average $W$ polarization. The full polarization information is contained 
in the 5-fold production and decay angular 
distributions,\cite{hpzh,lep2ww,lep2wwtgv}
\bq
{d^5\sigma\over d\cos\theta\;d\cos\theta_1^*\;d\phi_1^*\;
                             d\cos\theta_2^*\;d\phi_2^* }\; ,
\eq
Here the decay angles are defined in the $W^-$ and $W^+$ rest frames, 
respectively, and are measured against the direction of the parent $W$ in 
the lab. 

A major reason to study this 5-fold angular 
distribution is the experimental determination of the $WWZ$ and 
$WW\gamma$ couplings which enter in the first two Feynman graphs of 
Fig.~\ref{fig:feynman.ww}. This is analogous to the measurement of vector and 
axial vector couplings of the various fermions to the $Z$ and the $W$. 
The experiments at LEP2 so far confirm the SM predictions for these 
triple-gauge-boson couplings at about the 10\% level.\cite{lep2wwtgv,lep2ac}

Why would one consider $W$ decay angular distributions if one does not
want to measure polarizations or triple-gauge-boson couplings? As we saw
when discussing $W$ production, the produced $W^-$ is very strongly left-handed
polarized in the forward direction. This polarization has important 
consequences for the energy distribution of the decay products, and therefore
for the way the event appears in the detector. To be definite, let us 
consider the decay $W^-\to\ell^-\bar\nu$. In the $W$ rest frame, the 
four-momentum of the charged lepton is given by
\bq
p^*={m_W\over 2}\left(1,\sin\theta_1^*\cos\phi_1^*,\sin\theta_1^*\sin\phi_1^*,
\cos\theta_1^*\right)\;.
\eq
The charged lepton energy in the lab frame is obtained from here via a 
boost of its four-momentum, with a $\gamma$-factor 
$\gamma=E_W/m_W=\sqrt{s}/2m_W$,
\bq\label{eq:elWdecay}
p^0 = \gamma\bigl(p^{0*}+\beta p_z^*\bigr) = 
\gamma {m_W\over 2}\bigl(1+\beta\cos\theta_1^*\bigr) = 
{\sqrt{s}\over 4}\bigl(1+\beta\cos\theta_1^*\bigr)\; .
\eq
Thus, the polar angle of the lepton in the $W$ rest frame can be measured 
in terms of the lepton energy in the lab frame, and the two observables 
directly correspond to each other. This also implies that the energy 
distributions of the leptons in the lab are determined by their angular 
distributions in the $W$ rest frame, and these are fixed by the polarization
of the parent $W$.

As a concrete example, consider the average energy of the charged lepton, 
for the decay of a left-handed $W^-$. We need to average the result of 
Eq.~(\ref{eq:elWdecay}) over the normalized decay angular distribution, 
$3/8(1+\cos\theta^*)^2$, for a left-handed $W^-$:
\bq
\left< p^0(\ell^-) \right > ={E_W\over 2}
\int^1_{-1} d\cos\theta^* {3\over 8}(1+\cos\theta^*)^2 (1+\beta\cos\theta^*) 
= {E_W\over 2} (1+{\beta\over 2})\; .
\eq
Energy conservation fixes the average neutrino momentum to 
\bq
\left< p^0(\nu) \right > = {E_W\over 2} (1-{\beta\over 2})\; .
\eq
In the relativistic limit, $\beta\to 1$, the neutrino receives only 1/3 of the
energy of the charged lepton, on average, which has important consequences for 
detection and energy measurement of the leptons as well as for the 
consideration of $W$ pair production as backgrounds to new physics searches.

Polarization effects can have dramatic effects and one therefore needs 
predictions for $W$ pair production and decay which consider the full
chain $e^+e^-\to W^+W^-\to {4\;\rm fermions}$. In this full $2\to 4$ process
the $W$s merely appear as resonant propagators, which are treated as
Breit-Wigner resonances. Away from the peak of the resonance, seven additional
Feynman graphs contribute, beyond the three shown in Fig.~\ref{fig:feynman.ww},
even for the simplest case, $e^+e^-\to \mu^-\bar\nu_\mu\;u\bar d$. These
calculations have been performed~\cite{lep2ww} 
and are being used in the actual data analysis.

Another application, which nicely demonstrates the advantages of $\epem$ 
collisions, is the $W$-mass measurement in $\wpwm$ production at 
LEP2.\cite{LEP2wmass} Let 
us consider the decay $\wpwm\to \ell\nu jj$ as an example. A full 
reconstruction of the Breit-Wigner resonances, and a measurement of its center,
at $m_W$, is possible with the two jet momenta. However, the measurements of 
the jets' energies have large errors, of order $\pm 15\%$, and such a direct 
approach would lead to fairly large errors on the extracted $W$-mass. One can 
do much better by making use of the known kinematics of the event. 

\begin{figure}[t]
\centering\leavevmode
\psfig{figure=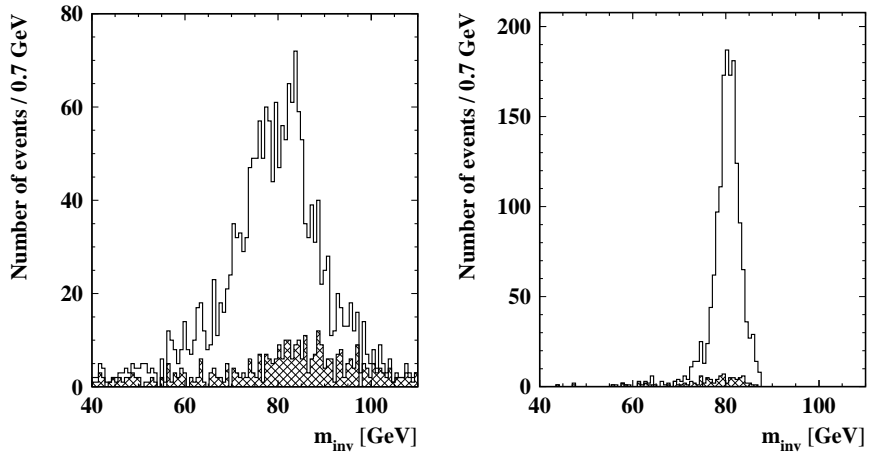,height=2.4in}
\vspace*{-0.1in}
\caption{
Expected $W$-mass reconstruction in $\wpwm\to\ell\nu jj$ events at LEP2.
The two plots show the invariant mass distribution of the $W$ decay products
before and after using the kinematic constraints described in the text.
From Ref.~\protect\cite{LEP2wmass}.
\label{fig:wmass.2C}
}
\vspace*{-0.1in}
\end{figure}

The 3-momentum of the neutrino in the event can be reconstructed from momentum 
conservation, as
\bq
{\bf p}_\nu = -{\bf p}_\ell - {\bf p}_{j_1} - {\bf p}_{j_2}\; ,
\eq
where we have used the fact that the lab frame is the c.m. frame, i.e. the sum
of all the final state momenta in the lab must add up to zero.
The energy of the massless neutrino is then given by 
$E_\nu=|{\bf p}_\nu |$.
Energy conservation and the equal masses of the two $W$s now imply the 
constraints
\bq
 E_{W_1} = E_{j_1} + E_{j_2} = E_b\; , \qquad
 E_{W_2} = E_{\ell} + E_{\nu} = E_b\; , 
\label{eq:beam.constr}
\eq
where $E_b=\sqrt{s}/2$ is the beam energy.
Even when considering the finite widths of the $W$ resonances and the 
possibility of initial state radiation, i.e. emission of photons along the beam
direction, which effectively lowers the c.m. energy $\sqrt{s}$, the constraint
of Eq.~(\ref{eq:beam.constr}) is satisfied to much higher accuracy than the 
precision of the jet energy measurement. One can thus drastically improve 
the $W$-mass resolution by using the two constraints to solve for the two
unknowns $E_{j_1}$ and  $E_{j_2}$ and use these values to calculate the 
$W^+$ and $W^-$ mass. The expected improvement is illustrated 
in Fig.~\ref{fig:wmass.2C}. 

First measurements of the $W$-mass, using both $\ell\nu jj$ and 4-jet events,
have already been performed with the 172 and 183~GeV data and resulted 
in~\cite{lep.mw} 
\bq
m_W=80.36 \pm 0.09 \; {\rm GeV}\;,
\eq
where the results from all four LEP experiments have been combined. Further
improvements are expected in the near future from the four times larger event
sample already collected in 1998. The LEP value agrees well with the one 
extracted from order $10^5$ leptonic $W$-decays observed at the Tevatron,
$m_W=80.41\pm 0.09$~GeV.\cite{tev.mw} 

\subsection{Chargino pair production} 
\label{sec:ee.chargino}

So far we have considered $\wpwm$ production as a signal. However, $W$-pairs
can be a serious background in the search for other new particles. Let us
consider one example in some detail, the production of charginos at LEP2.

\begin{figure}
\centering\leavevmode
\psfig{figure=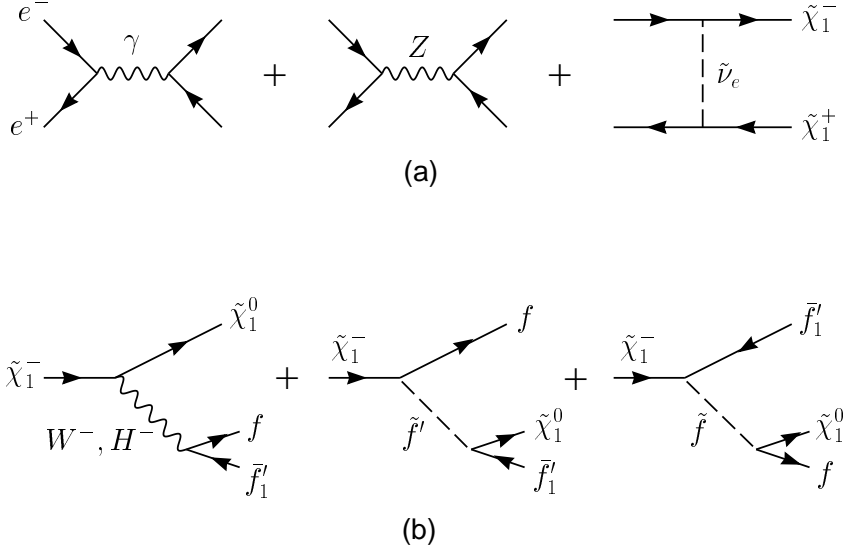,height=3.0in}
\vspace*{-0.1in}
\caption{
Feynman graphs for (a) chargino production in $\epem$ annihilation and 
(b) chargino decay into the lightest neutralino, $\tilde\chi^0_1$ and a pair 
of quarks or leptons.
\label{fig:feyn.chargino}
}
\vspace*{-0.1in}
\end{figure}

Charginos arise in supersymmetric models~\cite{mssm} 
as the fermionic partners of the
$W^\pm$ and the charged Higgs, $H^\pm$. Since they carry electric and weak 
charges, they couple to both the photon and the $Z$ and can be pair-produced 
in $\epem$ annihilation. The relevant Feynman graphs are shown in 
Fig.~\ref{fig:feyn.chargino}(a). Note that the graphs for chargino 
production are
completely analogous to the ones in Fig.~\ref{fig:feynman.ww} for $W$-pair
production, a reflection of the supersymmetry of the couplings. 

Two charginos are predicted by supersymmetry, and the lighter one, 
$\tilde\chi^\pm_1$, might be light enough to be pair-produced at 
LEP2.\cite{lep2.susy}
Production cross sections can be sizable, ranging from 2 to 5~pb at LEP2 for
typical parameters of SUSY models. However, the $s$-channel $\gamma$ and $Z$
exchange graphs, and $t$-channel sneutrino exchange in 
Fig.~\ref{fig:feyn.chargino}(a), interfere destructively. For small sneutrino 
masses, of order 100~GeV or less, this can lead to a drastic reduction in the
chargino pair production cross section, in particular near threshold. As a 
result, one must be prepared for production cross sections well below
1~pb at LEP2 also. In any case, the expected chargino pair cross section
can be smaller than the cross section for the dominant background, 
$\sigma(\epem\to\wpwm)\approx 17$~pb, by one order of magnitude or more.

A chargino, once it is produced, is expected to decay into the lightest 
neutralino, $\tilde\chi_1^0$, and known quarks and leptons, as shown in 
Fig.~\ref{fig:feyn.chargino}(b). The lightest neutralino is 
stable in most SUSY models, due to conserved $R$-parity, and does not 
interact inside the detector, thus leading to a missing
momentum signature. Over large regions of parameter space, where the squarks 
and sleptons entering the chargino decay graphs are quite heavy, the chargino
effectively decays into the neutralino and a virtual $W$. This results in 
a signature which is quite similar to $W$-pair production, since only the 
$W^*$ decay products are seen inside the detector. Thus, LEP looks for the 
charginos via the production and decay chain
\bq
\epem \to \tilde\chi_1^+\tilde\chi_1^- \to \tilde\chi_1^0\tilde\chi_1^0
W^{+*}W^{-*}\; ,
\eq
which, just like $\wpwm$ production, leads to a $jjjj$ signature 46\% of 
the time and to an $\ell\nu jj$ final state (including $\tau$'s) in 43.5\%
of all cases. The only difference is the additional presence of two massive
neutralinos, which needs to be exploited. 

Since the neutralinos escape detection, the energy deposited in the detector
typically is less than for $\wpwm$ events. In addition, the missing 
neutralinos spoil the momentum balance of the visible particles, which 
leads to a more spherically symmetric event than encountered in $W$ pair 
production. The most effective cut arises from the fact, 
however, that the four-momentum of the missing neutrinos and neutralinos 
can be reconstructed due to the beam constraint. Four-momentum conservation 
for the $\ell\nu jj\tilde\chi^0\tilde\chi^0$ final state state reads
\bq
p_{e^-}+p_{e^+}=p_\ell+p_{j_1}+p_{j_2}+\sla{p}\;
\eq
from which the missing mass, $M_{\rm miss}^2=\sla{p}^2$ can be reconstructed.
For a $\wpwm$ event the missing momentum corresponds to a single neutrino 
and, thus, $M_{\rm miss}=0$. For the signal, however, one has
\bq
\sla{p}^2 > 4 m_{\tilde\chi^0}^2\;.
\eq
Not only does the consideration of the missing mass allow for effective 
background reduction, to a level below 0.05--0.1~pb,\cite{lep2.susy} but 
it would also provide a measurement of the lightest neutralino mass, 
once charginos are discovered.

So far, no chargino signal has been observed at LEP. This pushes the chargino
mass bound above 90~GeV, provided the chargino-neutralino mass difference
is sufficiently large to allow enough energy for the visible chargino decay 
products.\cite{lep2.susyres} 
The LEP experiments have searched for other super-partners also, like
squarks and sleptons, and have not yet discovered any signals. The sfermions, 
since they are scalars, have a softer threshold turn-on, proportional to
$\beta^3$, and hence have a very low pair production cross section if their
mass is close to the beam energy. As a result, squark and slepton mass bounds 
currently are somewhat weaker than for charginos, but even here, sfermions with
masses below 70--85~GeV (depending on flavor) are excluded.\cite{lep2.susyres}

\subsection{Future $\epem$ and $\mu^+\mu^-$ colliders}
\label{sec:ee.mumu}

An exciting search presently being conducted at LEP is the 
hunt for the Higgs boson, in $\epem\to ZH$.\cite{lep2H} 
The mass of the Higgs boson 
does influence radiative corrections to 4-fermion amplitudes, via
$ZH$ and $WH$ loops contributing to the $Z$ and $W$ propagator corrections.
Precise measurements of asymmetries in $\epem\to\bar ff$, of partial $Z$ widths
to leptons and quarks, of atomic parity violation etc. allow to extract
the expected Higgs mass within the SM. These measurements point to a relatively
small Higgs mass, of about 100~GeV, albeit with a large error of about a 
factor of two.\cite{altar,langack,marciano} LEP is exactly searching in 
this region.

In $\epem\to ZH$, the large mass of the accompanying $Z$ limits the reach 
of the LEP experiments, to about~\cite{lep2H}
\bq
m_H \lsim \sqrt{s}-m_Z-5\;{\rm GeV}\; .
\label{eq:mHlim.ee}
\eq
With an  eventual c.m. energy of $\sqrt{s}\approx 200$~GeV, 
this allows discovery of the Higgs at LEP, provided its mass is below 
about 105~GeV. However, measurements at energies up to 189~GeV have not 
discovered anything yet, setting a lower Higgs mass bound, within the SM, 
of 95~GeV.\cite{marciano} We need luck to still find a Higgs signal at LEP, 
before the LEP tunnel needs to be cleared for installing the LHC.

Given the indications for a relatively light Higgs from electroweak
precision data, an expectation which is shared by supersymmetric 
models,\cite{mssm} an $\epem$ collider with higher c.m. energy 
than LEP2 is called for. As explained in Section~\ref{sec:overview} this
cannot be a circular machine, due to excessive synchrotron radiation,
but rather should be a linear $\epem$ collider.\cite{NLC}
A 500~GeV NLC, with a yearly
integrated luminosity of 10--100~fb$^{-1}$,  would be a veritable Higgs 
factory. At such a machine,  the Higgs production cross section is of 
order 0.1~pb in both
the $\epem\to ZH$ production channel (for $m_H\lsim 300$~GeV) and also
in the weak boson fusion channel, where the Higgs 
boson is radiated off a $t$-channel $W$ 
($\sigma(\epem\to H\nu\bar\nu)\gsim 0.03$~pb for $m_H\lsim 200$~GeV). 
In this mass range, the 1000 to 10000 produced Higgs bosons per year 
would allow for detailed investigations of Higgs boson properties, in the 
clean environment of an $\epem$ collider. Somewhat higher Higgs boson masses,
up to $m_H\approx \sqrt{s}-100$~GeV (see Eq.~(\ref{eq:mHlim.ee})) are 
accessible as well, albeit with lower production rates.

The NLC would greatly extend the search region for other new heavy particles
as well, like the charginos and neutralinos of the MSSM, or its squarks
and sleptons. Even if theses particles are first discovered at a hadron 
collider like the LHC, the cleaner environment of $\epem$ collisions, the more
constrained kinematics, and the observability of most of the decay channels
give linear $\epem$ colliders great advantages for detailed studies of the
properties of any new particles.

This is true also for the latest new particle that has been discovered 
already, the top quark.\cite{topCDFprl,topD0} A scan of the top production
threshold in $\epem$ collisions, at $\sqrt{s}=2m_t\approx 350$~GeV, would 
give an unprecedented precision in the measurement of the top quark mass.
The simultaneous direct measurement of the top quark width would determine
the $V_{tb}$ CKM matrix element and thus provide a significant test of the
electroweak sector.

All these $\epem$ collider measurements could also be performed at
a $\mu^+\mu^-$ collider.\cite{mupmum}
Such a machine would have the added advantage of
an excellent beam energy resolution, of order $10^{-4}$ or better, while
beam-strahlung in the tight focus of an $\epem$ linear collider leads to 
a significant smearing of the c.m. energy. The very precisely determined 
beam energy can then be used for a scan of the $\bar tt$ 
production threshold, which resolves detailed features like the location 
of the (extremely short-lived) first $\bar tt$ bound state, QCD binding
effects and the value of the strong coupling constant, or even Higgs exchange
effects on the shape of the $\bar tt$ production threshold.

Another advantage of a $\mu^+\mu^-$ collider is the larger coupling of the 
Higgs boson to the muon as compared to the electron, due to the muon's 200
times larger mass. This allows the direct $s$-channel production of the Higgs
resonance in muon collisions, $\mu^+\mu^-\to H$. Because of the excellent 
energy resolution of a muon collider, an energy scan of the Higgs resonance 
would provide us with a very precise measurement of the Higgs boson mass,
with an error of order MeV, and if dedicated efforts are made to keep the
energy spread as small as possible, even the full width of the Higgs resonance
can be determined directly.\cite{mupmum}

These examples clearly show that a linear $\epem$ collider or a muon collider
would be a terrific experimental tool and would greatly advance our 
understanding of particle interactions. Unfortunately, no such machine has 
been approved for construction yet. And it may be argued that we first 
need to establish the existence of new heavy particles before investing 
several billion dollars or euros into a machine to search for them and then
study their detailed 
properties. For many of the particles predicted by supersymmetry, or for 
the Higgs boson, the machines needed for discovery already exist or are
under construction, namely the Tevatron at Fermilab and the LHC at CERN.

\section{Hadron Colliders}
\label{sec:pp}

The highest center of mass energies and, hence, the best reach for new heavy 
particles is provided by hadron colliders, the Tevatron with its 2~TeV
$\bar pp$ collisions at present, and the LHC with 14~TeV $pp$ collisions
after 2005. At the Tevatron the top quark has been discovered in 1994, and 
Higgs and supersymmetry searches will resume in run II. The LHC is expected
to do detailed investigations of the Higgs sector, and should answer
the question whether TeV scale supersymmetry is realized in nature. 
Before discussing how these studies can be performed in a hadron collider
environment, we need to consider the general properties of production
processes at these machines in some detail.

\subsection{Hadrons and partons}
\label{sec:intro.hadrons}

\begin{figure}
\centering\leavevmode
\psfig{figure=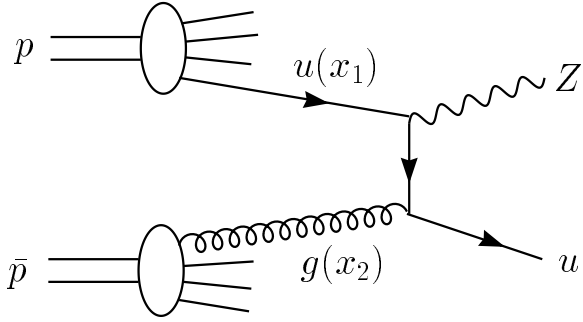,height=1.9in}
\caption{
One of the subprocesses contributing to $Zj$ production in $p\bar p$ 
collisions. The two initial partons carry momentum fractions $x_1$ and 
$x_2$ of the incoming  proton and anti-proton, respectively.
\label{fig:ppzj}
}
\vspace*{-0.1in}
\end{figure}

A typical hard hadronic collision is sketched in 
Fig.~\ref{fig:ppzj}: one of the subprocesses contributing to $Z+$jet 
production, namely $ug\to uZ$. The up-quark and the gluon carry a fraction
of the parent proton momenta only, $x_1$ and $x_2$, respectively. Thus, the 
incoming parton momenta are given by
\ba
p_u & = & x_1 p = x_1 {\sqrt{s}\over 2} \bigl( 1,0,0,1)\; ,\nonumber \\
p_g & = & x_2 \bar p = x_2 {\sqrt{s}\over 2} \bigl( 1,0,0,-1)\; ,
\label{eq:p_parton}
\ea
and the available center of mass energy for the $Z+$jet final state is given
by the root of $\hat s = (p_u+p_g)^2=2p_u\cdot p_g = x_1x_2s$, i.e. it is only
a fraction $\sqrt{x_1x_2}$ of the collider energy.

In order to calculate observable production rates, for a process 
$p\bar p\to X$, we first need to identify all the parton level subprocesses
$a_1+a_2\to \hat x=b_1+b_2+\cdots+ b_n$ which give the desired signature. 
In the above example of $Z+$jet production this includes, at tree level, 
$u+\bar u\to Z+g$, $\bar u+u\to Z+g$ (i.e. the subprocess where the anti-$u$ 
comes from the proton), $g+u\to Z+u$, $u+g\to Z+u$, $g+\bar u\to Z+\bar u$, 
$\bar u+g\to Z+\bar u$, and all
the corresponding subprocesses with the up quark replaced by down, strange, 
charm, and bottom quarks, which are all treated as massless partons inside
the proton. 
The full cross section for the process $p\bar p\to X$ is then given by
\ba
\sigma & = & \int dx_1dx_2 \sum_{\rm subprocesses}\; 
             f_{a_1/p}(x_1)\; f_{a_2/\bar p}(x_2)\; \nonumber \\  
&& {1\over 2\hat s}\int d\Phi_n(x_1p+x_2\bar p;\;p_1\dots p_n) 
   \Theta({\rm cuts}) \nonumber \\
&& \overline{\sum}|{\cal M}|^2(a_1a_2\to b_1b_2\dots b_n)\; .
\label{eq:sigma}
\ea
Here $f_{a_1/p}(x_1)$ is the probability to find parton $a_1$ inside the 
proton, carrying a fraction $x_1$ of the proton momentum, i.e. the $a_1$
parton distribution function. Similarly, $f_{a_2/\bar p}(x_2)$ is the $a_2$ 
parton distribution function (pdf) inside the anti-proton. In the second line 
of (\ref{eq:sigma}) $1/2\hat s$ is the flux factor for the partonic cross 
section, 
\bq
d\Phi_n(P;\;p_1\dots p_n) = 
\prod_{i=1}^n \left( {d^3{\bf p}_i\over (2\pi)^3 2E_i} \right)
(2\pi)^4\;\delta^4(P-\sum_i p_i)
\eq
is the Lorentz invariant phase space element, and $\Theta({\rm cuts})$ 
is the acceptance function, which summarizes the kinematical cuts on all the 
final state particles, i.e. $\Theta = 1$ if all the partons $a_i$ satisfy all 
the acceptance cuts and $\Theta=0$ otherwise. Finally, ${\cal M}$, in the
third line of (\ref{eq:sigma}) is the Feynman amplitude for the subprocess 
in question, squared and summed/averaged over the polarizations and colors 
of the external partons. 

Compared to the calculation of cross sections for $\epem$ collision, the 
new features are integration over pdf's and the fact that a much larger
number of partonic subprocesses must be considered for a given experimental
signature. The introduction of pdf's introduces an additional uncertainty
since they need to be extracted from other data, like deep inelastic 
scattering, $W$ production at the Tevatron, or direct photon production.
The extraction of pdf's is continuously being updated and refined, and in 
practice, hadron collider cross sections are calculated by using numerical
interpolations which are provided by the groups improving the pdf 
sets.\cite{pdf} These pdf determinations have been dramatically improved 
over time and typical pdf uncertainties now are below the 5\% 
level, at least in the range $10^{-3}\lsim x\lsim 0.2$, which is 
most important for our discussion of new particle production processes.

More important than the appearance of pdf's are the kinematic effects 
which result from the fact that the hard collision is between partons.
Since momenta of the incoming partons are not known a priori, we cannot 
make use of a beam energy constraint as in the case of $\epem$ collisions. 
The missing information on the momentum parallel to the beam axis affects 
the analysis of events with unobserved particles in the final state, 
like neutrinos or the lightest neutralino. Momentum conservation 
can only be used for the components transverse to the beam axis, i.e.
only the missing transverse momentum vector, ${\sla{\bf p}}_T$, can be
reconstructed. 

Another effect is that the lab frame and the c.m. frame of the hard collision
no longer coincide. Rather the partonic c.m. system receives a longitudinal 
boost in the direction of the beam axis, which depends
on $x_1/x_2$. This longitudinal boost is most easily taken into account 
by describing four-momenta in terms of rapidity $y$ instead of scattering 
angle $\theta$. For a momentum vector
\bq
p = (E,p_x,p_y,p_z) = 
E(1,\beta\sin\theta\cos\phi,\beta\sin\theta\sin\phi,\beta\cos\theta)\;,
\eq
rapidity is defined as 
\bq
y={1\over 2}\log{E+p_z\over E-p_z}\; ,
\eq
which, in the massless limit ($\beta\to 1$), reduces to pseudo-rapidity,
\bq
\eta= {1\over 2}\log{1+\cos\theta\over 1-\cos\theta}\; .
\eq
The advantage of using rapidity is that under an arbitrary boost along the
$z$-axis, rapidity differences remain invariant, i.e. they directly 
measure relative scattering angles in the partonic c.m. frame. Using  
Eq.~(\ref{eq:p_parton}), the c.m. momentum is given by 
$P=\sqrt{s}/2(x_1+x_2,0,0,x_1-x_2)$ which results in a c.m. rapidity
$y_{c.m.}=1/2\log(x_1/x_2)$. As a result, rapidities $y^*$ in the partonic c.m.
system and rapidities $y$ in the lab frame are connected by
\bq
y = y^*+y_{c.m.} = y^*+{1\over 2}\log{x_1\over x_2}\; ,
\eq
a relation which can be used to determine all scattering angles in the 
theoretically simpler partonic rest frame whenever all final state momenta
can be measured and, hence, the c.m. momentum $P$ is known.

\subsection{$Z$ and $W$ Production}
\label{sec:drell.yan}

One of the early highlights of $\bar pp$ colliders was the discovery of the 
$W$ and $Z$ bosons at the CERN S$\bar pp$S, a 630~GeV $\bar pp$ 
collider.\cite{wzdisc} $W$ and $Z$ production provide a nice example which
demonstrates the use of some of the transverse
observables discussed in the previous subsection. In addition, the story might
repeat, since nature might have an additional neutral or charged heavy gauge 
boson in store, a $Z'$ or an $W'$ which might appear at the LHC. Let us thus
consider $Z$ production in some detail.

The prototypical Drell-Yan process is $\bar pp\to Z\to \ell^+\ell^-$, where,
to lowest order in the strong coupling constant, the $Z$ can be produced by 
annihilation of a quark-antiquark pair . The partonic subprocess 
$\bar qq\to \ell^+\ell^-$ leads to two leptons with balancing transverse
momenta, which can be parameterized in terms of their lab frame $p_T$,
pseudo-rapidity $\eta$, and azimuthal angle $\phi$,
\ba\label{eq:pl.lab}
\ell &=& p_T\bigl( \cosh\eta,\cos\phi,\sin\phi,\sinh\eta\bigr)\; , \\
\bar\ell &=& p_T\bigl( \cosh\bar\eta,-\cos\phi,-\sin\phi,\sinh\bar\eta\bigr)\;.
\ea
Since transverse momentum is invariant under a boost along the $z$-axis, we
may as well determine it in the partonic c.m. frame, where the $\ell^-$ 
momentum is given by
\bq\label{eq:pl.cm}
\ell^*={m_Z\over 2}\bigl(1,
\sin\theta^*\cos\phi,\sin\theta^*\sin\phi,\cos\theta^*\bigr)\; .
\eq
One finds $p_T=m_Z/2\;\sin\theta^*$  by equating the transverse momenta 
in the two frames. This implies
\bq
|\cos\theta^*| = \sqrt{1-\sin^2\theta^*} = \sqrt{ 1-{4p_T^2\over m_Z^2}}\;.
\eq
Using this relation we obtain the transverse momentum spectrum in terms of the
lepton angular distribution in the c.m. frame,
\bq\label{eq:jacob}
{d\sigma\over dp_T^2} = {d\sigma\over d\cos\theta^*} 
\left|{d\cos\theta*\over dp_T^2}\right| = {d\sigma\over d\cos\theta^*} 
{1\over m_Z\sqrt{m_Z^2/4 - p_T^2}}\; .
\eq
The $p_T$ distribution diverges at the maximum transverse momentum 
value, $p_T=m_Z/2$.
This Jacobian-peak, so called because it arises from the Jacobian factor
in Eq.~(\ref{eq:jacob}), is smeared out in practice by finite detector 
resolution, the finite $Z$-width and QCD effects. Nevertheless, it is an
excellent tool to determine the $W$ mass in the analogous $W\to\ell\nu$
decay, which has the Jacobian peak at half the $W$ mass. 

\begin{figure}
\centering\leavevmode
\psfig{figure=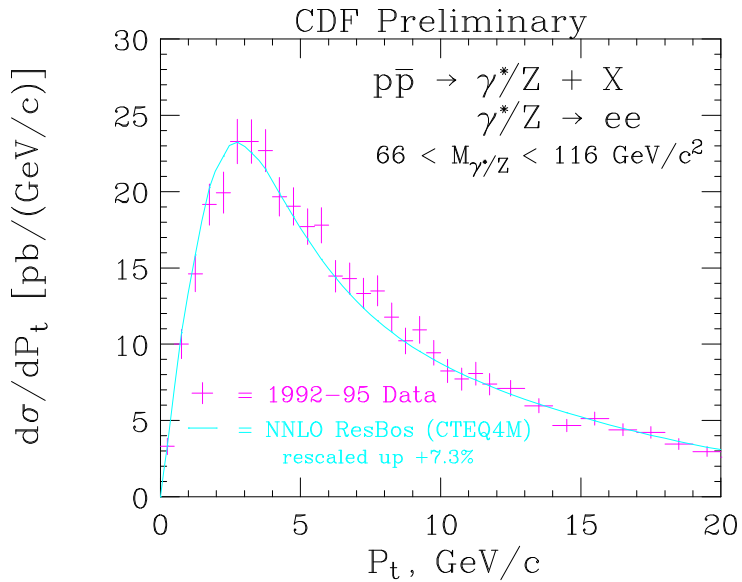,height=3.0in}
\vspace*{-0.1in}
\caption{
Transverse momentum distribution of the produced lepton pair in Drell-Yan 
production, $\bar qq\to \gamma^*/Z\to \epem$, as measured by the CDF 
Collaboration at the Tevatron.
\label{fig:ptz}
}
\vspace*{-0.1in}
\end{figure}

The Jacobian peak is smeared out considerably by QCD effects, namely the 
emission of additional partons in Drell-Yan production. Only at lowest order,
in $\bar qq\to \ell^+\ell^-$ or $\bar qq'\to\ell\nu$, do the transverse 
momenta of the two decay leptons exactly balance. Taking QCD effects
into account, we must consider gluon radiation or subprocesses like $qg\to qZ$
as depicted in Fig.~\ref{fig:ppzj}. The lepton pair now obtains a transverse
momentum, which balances the transverse momentum of the additional 
parton(s) in the final state. In fact, multiple soft gluon emission renders
a zero probability to lepton pairs with $p_T(Z)=0$. In real life, their 
transverse momentum distribution peaks at a few GeV, as can be seen in 
Fig.~\ref{fig:ptz}, which shows the $p_T(\epem)$ distribution as observed 
at the Tevatron.

When trying to determine the kinematics of an event to better than some
10\% ($={\cal O}(\alpha_s)$), we need to take soft parton emission into 
account. One way to do this is to use transverse mass instead of transverse
momentum of the lepton. For $W\to\ell\nu$ decay the transverse mass is 
defined as
\bq
m_T(\ell,{\sla{\bf p}}_T) = \sqrt{ (E_{T\ell}+{\sla E}_T)^2 - 
({\bf p}_{T\ell}+{\sla{\bf p}}_T)^2 }\; ,
\eq
i.e. it is determined from the difference of squares of the transverse energy 
and the transverse momentum vector of the $\ell\nu$ pair. This is 
analogous to the 
definition of invariant mass, which in addition includes the contributions 
from the longitudinal momentum, along the beam axis. The transverse mass 
retains a Jacobian peak, at $m_T(\ell,{\sla{\bf p}}_T)=m_W$, even in the
presence of QCD radiation.

\subsection{Extra $W$ and $Z$ bosons}
\label{sec:extraWZ}

The SM is a gauge theory based on the gauge group $G_{SM}=SU(3)\times 
SU(2)_L\times U(1)_Y$ and each of these factors is associated with a set 
of gauge bosons, the eight gluons of $SU(3)$, the photon and the $W^\pm$ 
and $Z$ of the electroweak sector. It is possible, however, that the gauge 
symmetry of nature is larger, which in turn would predict the existence of 
extra gauge bosons. The apparent symmetry of the SM would arise because the
extra gauge bosons are too heavy to have been observed as yet, made massive
by the spontaneous breaking of the extra gauge symmetry. Examples of such 
extended gauge sectors are left-right symmetric models~\cite{wprime} with
\bq
G=SU(3)\times SU(2)_R\times SU(2)_L\times U(1)_{B-L}\;,
\eq
where the new $SU(2)_R$ factor gives rise to an extra charged $W'$ with $V+A$
couplings to quarks and leptons and an additional $Z'$, or extensions with
extra $U(1)$ factors,\cite{cvetic,zprime}
\bq
G=SU(3)\times SU(2)_L\times U(1)\times U(1)\;,
\eq
which lead to the existence of an extra $Z'$. Present
indirect limits on such extra gauge bosons are relatively weak, and allow 
extra $W'$ are $Z'$ bosons to exist with masses above some 500~GeV.\cite{chm}
These indirect bounds are obtained form the apparent absence of additional
contact interactions, similar to Fermi's four-fermion couplings, in low energy
data. Given mass bounds of a few hundred GeV only, additional gauge bosons, 
with masses up to several TeV, could readily be observed at the LHC.

The cleanest method for discovering extra $Z'$ bosons would be through a 
repetition of the historic CERN experiments~\cite{wzdisc} which lead to 
the discovery of the $Z$ in 1983, i.e. by searching for the $Z'$ resonance 
peak in
\bq
\bar qq \to Z' \to \ell^+\ell^-\;, \qquad \ell=e,\mu\; .
\eq
The reach of this search depends on the coupling $g'^f_{L,R}$ of the extra 
$Z'$ to left- and right-handed fermions. The production cross section is 
proportional to $(g'^q_L)^2+(g'^q_R)^2$. The decay branching fraction,
$B(Z'\to\ell^+\ell^-)$,  depends on the relative size of this combination 
of left- and right-handed couplings for lepton $\ell$ to the same combination,
summed over all fermions. If the product of production
cross section times leptonic branching ratio,
\bq
\sigma\cdot B = \sigma(pp\to Z')\;B(Z'\to\ell^+\ell^-)\; ,
\eq
is the same as for the SM $Z$-boson, scaled up to $Z'$ mass, i.e. if the 
couplings of the $Z'$ are SM-like, the LHC experiments can observe a $Z'$ 
with a mass up to $m_{Z'}=5$~TeV. Smaller (larger) couplings would decrease 
(increase) this reach, of course.\cite{ATLAS}

While the LHC will not be capable of repeating the precision experiments of
LEP/SLC, for a $Z'$, a lot of additional information can be obtained by more 
detailed observations of leptonic $Z'$ decays.\cite{zprime} 
One measurement which would be
of particular importance is the determination of the lepton charge asymmetry.
At the parton level, the forward-backward charge asymmetry measures the 
relative number of $\bar qq\to Z'\to\ell^+\ell^-$ events where the $\ell^-$ 
goes into the same hemisphere as the incident quark, as compared to events 
where the $\ell^+$ goes into the quark direction. In terms of the 
pseudo-rapidity of the $\ell^-$, $\eta^*$, as measured relative to the 
incident quark direction, i.e.
\bq
\eta^*= {1\over 2}\log{1+\cos\theta^*\over 1-\cos\theta^*}\; ,
\eq
where $\theta^*$ is the c.m. frame angle between the incident quark and the 
final state $\ell^-$ (or the angle between the incident anti-quark and the 
final state $\ell^+$), the forward-backward asymmetry, at the parton level,
is given by
\bq
\hat A_{FB}^\ell = {\hat\sigma(\eta^*>0)-\hat\sigma(\eta^*<0)\over
\hat\sigma(\eta^*>0)+\hat\sigma(\eta^*<0)} = 
{3\over 4}\, {(g'^q_R)^2-(g'^q_L)^2\over (g'^q_R)^2+(g'^q_L)^2}\;
{(g'^{\ell}_R)^2-(g'^{\ell}_L)^2\over (g'^{\ell}_R)^2+(g'^{\ell}_L)^2}
\eq
One sees that a measurement of the forward-backward asymmetry gives a direct
comparison of left-handed and right-handed couplings of the $Z'$ to leptons 
and quarks.

Unfortunately, at a $pp$-collider, the two proton beams have equal 
probabilities
to originate the  quark or the antiquark in the collision, and, therefore, 
the forward backward asymmetry averages to zero when considering all events.
One can make use of the different Feynman $x$ distributions of up and down 
quarks as opposed to anti-quarks, however. At small $x$, quarks and anti-quarks
have roughly equal pdf's, $q(x)\approx \bar q(x)$, while at large $x$ the 
valence quarks dominate by a sizable fraction, $q(x)\gg \bar q(x)$. 
In the experiment one measures the rapidities $y_+$ and $y_-$ of the $\ell^+$
and $\ell^-$, respectively. Since the leptons are back-to-back in the c.m.
frame, their c.m. rapidities cancel, and the sum of the lab frame rapidities
gives the rapidity $y=y_{c.m.}$ of the c.m. frame,
\bq
y = {1\over 2}(y_+ + y_-) = {1\over 2}\log{x_1\over x_2}\; ,
\eq
while their difference measures $\eta^*$,
\bq
\eta^*= {1\over 2}\log{1+\cos\theta^*\over 1-\cos\theta^*}
={1\over 2}(y_- - y_+)\; .
\eq
At $y>0$ we have $x_1>x_2$, and therefore it is more likely that the 
quark came from the left, while at $y<0$ an anti-quark from the left 
is dominant. Measuring both $\eta^*$ and $y$, the charge asymmetry, 
\bq
A(y) = {{d\sigma\over dy}(\eta^*>0)-{d\sigma\over dy}(\eta^*<0)\over
{d\sigma\over dy}(\eta^*>0)+{d\sigma\over dy}(\eta^*<0)} \; ,
\eq
can be determined. Of course, we have $A(y)=-A(-y)$ at a $pp$ collider, where
the two sides are equivalent, and the average over all $y$ vanishes. At fixed 
$y$, however, $A(y)$ is analogous to the forward-backward charge asymmetry at
$\epem$ colliders, and it measures the relative size of left-handed and 
right-handed $Z'$ couplings.

Different extra gauge groups predict substantially different sizes for left-
and right-handed couplings of the $Z'$ to quarks and leptons. Thus, the
lepton charge asymmetry, $A(y)$, is a very powerful tool to distinguish 
different models, once a $Z'$ has been discovered.

Similar to the $Z'$ search, the search for a heavy charged gauge boson, $W'$,
would repeat the $W$ search at CERN in the early eighties. One would study
events consisting of a charged lepton and a neutrino, signified by missing
transverse momentum opposite to the charged lepton direction. The mass
of the $W'$ is then determined by the Jacobian peak in the transverse mass
distribution, at $m_T(\ell,{\sla{\bf p}}_T)=m_{W'}$. For a $W'$ with SM 
strength couplings, the LHC can find it and measure its mass, up to $W'$ 
masses of order 5~TeV, by searching for a shoulder in the transverse mass 
distribution and measuring its cutoff at $m_{W'}$.\cite{ATLAS}

\subsection{Top search at the Tevatron}
\label{sec:top}

The leptonic decay of a $W$ or $Z$ produces a fairly clean signature at a 
hadron collider. Perhaps more typical for a new particle search was the 
discovery of the top quark~\cite{topCDFprl,topD0} 
at the Fermilab Tevatron. A much more complex signal needed to be isolated 
from large QCD backgrounds. At the same time the top discovery provides a 
beautiful example for the use of hadronic 
jets as a tool for discovering new particles. Let us have a brief, historical
look at the top quark search at the Tevatron, from this particular viewpoint.

In $p\bar p$ collisions at the Tevatron, 
the top quark is produced via quark anti-quark annihilation, $q\bar q \to
t\bar t$, and, less importantly, via $gg\to t\bar t$. Production cross
sections have been calculated at next-to-leading order, and are
expected to be around 5~pb for a top mass of 175~GeV.\cite{ttNLO}
The large top decay width which is expected in the SM,
\bq
\Gamma (t\to W^+ b) \approx 1.6\;{\rm GeV}\;,
\eq
implies that the $t$ and $\bar t$ decay well before hadronization, and the 
same is true for the subsequent decay of the $W$ bosons. Thus, a parton 
level simulation for the complete decay chain, including final parton 
correlations, is a reliable means of predicting detailed properties of the 
signal. The top quark signal, $t\bar t\to bW^+\;\bar bW^-$, is determined by
the various decay modes of the $\wpwm$ pair, whose branching ratios were 
discussed in Sec.~\ref{sec:ee.ww}. In order to distinguish the signal
from multi-jet backgrounds, 
the leptonic decay $W\to\ell\nu$ ($\ell=e,\;\mu$) of at least one of the two 
final state $W$s is extremely helpful. On the other hand, the leptonic decay 
of both $W$s into electrons or muons has a branching ratio of 
$\approx 4\%$ only, and thus the prime top search channel is the 
decay chain
\bq
t\bar t\;\to\; bW^+\;\bar bW^-\;\to\; \ell^\pm\nu\; q\bar q\; b\bar b\; .
\label{eq:topsig}
\eq
Within the SM, this channel 
has an expected branching ratio of $\approx 30\%$. After 
hadronization each of the final state quarks in (\ref{eq:topsig}) may 
emerge as a hadronic jet, provided it carries enough energy. Thus the 
$t\bar t$ signal is expected in $W+3$~jet and $W+4$~jet 
events\footnote{Gluon bremsstrahlung may increase the number of jets further 
and thus all $W+\geq 3$~jet events are potential $t\bar t$ candidates.}.

\begin{figure}[t]
\centering\leavevmode
\psfig{figure=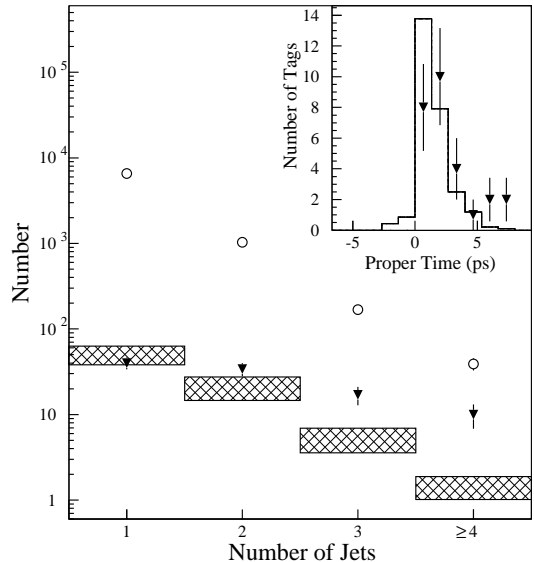,height=3.0in}
\vspace*{-0.1in}
\caption{
Number of $W+n$~jet events in the CDF top quark search as a function of jet 
multiplicity. Number of observed events are given without $b$-tagging 
(open circles) and with an SVX tag (triangles). The expected background, 
mainly from QCD $W+n$~jet events, is given by the cross-hatched bars.
From the original CDF top quark discovery paper, Ref.~\protect\cite{topCDFprl}.
\label{fig:cdftop1}
}
\vspace*{-0.1in}
\end{figure}

Events with leptonic $W$ decays and several jets can also arise from QCD 
corrections to the basic Drell-Yan process $q\bar q\to W^\pm \to \ell^\pm\nu$.
The process $ug\to dggW^+$, for example, will give rise to $W+3$~jet events
and its cross section and the cross sections for all other subprocesses with 
a $W$ and three partons in the final state need to be calculated in order 
to assess the QCD background of $W+3$~jet events, at tree level. $W+n$~jet 
cross sections have been calculated for $n=3$ jets~\cite{W3j} and $n=4$ 
jets.\cite{W4j} As in the experiment, the calculated
$W+n$~jet cross sections depend critically on the minimal transverse energy
of a jet. CDF, for example, requires a cluster of hadrons to carry 
$E_T>15$~GeV to be identified as a jet,\cite{topCDF} and this observed 
$E_T$ must then be translated into the corresponding parton transverse 
momentum in order to get a prediction for the $W+n$~jet cross sections. 

At this level the QCD backgrounds are still too large to give a viable top
quark signal. The situation was improved substantially by using the fact that 
two of the four final state partons in the signal are $b$-quarks, while only 
a small fraction of the $W+n$~parton background events have $b$-quarks in the 
final state. These fractions are readily calculated by using $W+n$~jet Monte 
Carlo programs. There are several experimental techniques
to identify $b$-quark jets, all based on the weak decays of the produced
$b$'s. One method is to use the finite $b$ lifetime of about $\tau=1.5$~ps 
which leads to $b$-decay vertices which are displaced by 
$\gamma c\tau=$~few mm from the primary interaction vertex. These 
displaced vertices can be resolved by precision tracking, with the aid of 
their Silicon VerteX detector in the case of CDF, and the method is, 
therefore, called  SVX tag. In a second method, $b$ decays are identified 
by the soft leptons which arise in the weak decay chain 
$b\to W^*c,\; c\to W^*s$, where either one of the virtual $W$s may decay 
leptonically.\cite{topD0,topCDF}

The combined results of using jet multiplicities and SVX $b$-tagging to
isolate the top quark signal are shown in Fig.~\ref{fig:cdftop1}. A clear
excess of $b$-tagged 3 and 4 jet events is observed above the expected 
background. The excess events would become insignificant if all jet 
multiplicities were combined or if no $b$-tag were used (see open circles).
Thus jet counting and the identification of $b$-quark jets are
critical for identification of top quark events.

\begin{figure}[t]
\centering\leavevmode
\psfig{figure=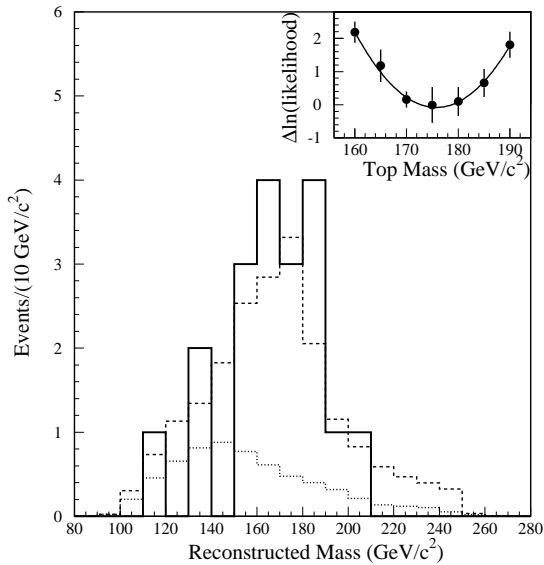,height=3.0in}
\vspace*{-0.1in}
\caption{
Reconstructed  mass distribution for $W+\geq 4$~jet events with a
$b$-tag. The solid histogram represents the CDF data. Also included are the
expected background (dotted histogram) and the expected signal+background
for $m_t = 175$~GeV. The insert shows the likelihood fit to determine the top 
quark mass, which yielded $m_t=176\pm 8\pm 10$~GeV at the time. From 
Ref.~\protect\cite{topCDFprl}.
\label{fig:cdfmt}
}
\vspace*{-0.1in}
\end{figure}

Beyond counting the number of jets above a certain transverse energy, the
more detailed kinematic distributions, their summed scalar 
$E_T$'s~\cite{topD0} and multi-jet invariant masses, have also been 
critical in the top quark search.
The top quark mass determination, for example, relies on a good understanding
of these distributions. Ideally, in a $t\bar t\to (\ell^+\nu b)(q\bar q\bar b)$
event, for example, the two subsystem invariant masses should be equal to the
top quark mass,
\bq 
m_t \approx m(\ell^+\nu b) \approx m(q\bar q\bar b)\; .
\eq
Including measurement errors, wrong assignment of observed jets to the two
clusters, etc. one needs to perform a constrained fit to extract $m_t$. The
1995 CDF result of this fit~\cite{topCDFprl} is shown in Fig.~\ref{fig:cdfmt}. 
In addition, the figure  demonstrates that the observed $b$-tagged
$W+4$~jet events (solid histogram) are considerably harder than the QCD
background (dotted histogram). On the other hand the data agree very well 
with the top quark hypothesis (dashed histogram). 

By now, the top quark has been observed in all three decay channels of the 
$\wpwm$ pair, purely leptonic $W$ decays, $\wpwm\to \ell\nu jj$, and 
4-jet decays, and all channels have been used to extract the top-quark mass.
Results at present are 
\bq 
m_t = 172.1 \pm 7.1\; {\rm GeV}\; ,
\eq
from the D0 Collaboration,\cite{D0mt} and
\bq 
m_t = 176.0 \pm 6.5\; {\rm GeV}\; ,
\eq
from the CDF Collaboration.\cite{CDFmt}

\section{Higgs search at the LHC}
\label{sec:Higgs.at.LHC}

With the top-quark discovery at the Tevatron, the elementary fermions 
of the SM have all been observed. The missing ingredient, as far as the SM
is concerned, is the Higgs boson. LEP2 is likely to find it if its mass is 
below $\approx 105$~GeV. The Tevatron has a chance to discover the Higgs 
boson in the processes $\bar qq\to WH,\; H\to\bar bb$ (for Higgs masses below
$\approx 130$~GeV)~\cite{kuhlmann} or $gg\to H\to WW^*$ (for 
130~GeV$\lsim m_H\lsim 180$~GeV)~\cite{han} if sufficient 
luminosity can be collected within the next few years. (Between 10 and 
30~fb$^{-1}$ are required for this purpose.) The best candidate for Higgs
discovery and detailed Higgs studies within the next ten years is 
the LHC, however.

\begin{figure}
\centering\leavevmode
\psfig{figure=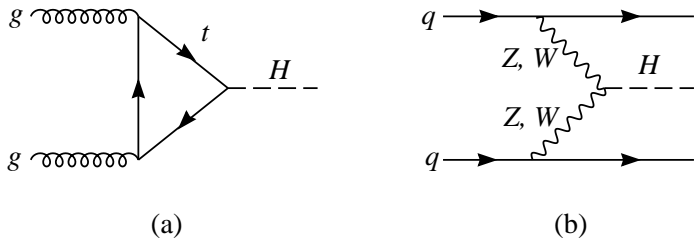,height=1.4in}
\vspace*{-0.1in}
\caption{
Feynman graphs for the two dominant Higgs production processes at the LHC,
(a) gluon-gluon fusion via a top-quark loop and (b) weak boson fusion.
\label{fig:Hprodfeyn}
}
\vspace*{-0.1in}
\end{figure}

\begin{figure}[t]
\centering\leavevmode
\psfig{figure=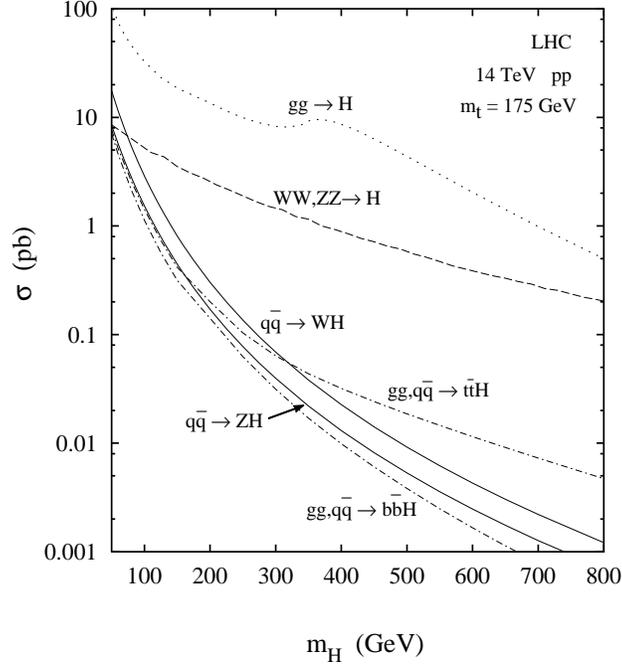,height=3.5in}
\vspace*{-0.2in}
\caption{
Production cross sections for the SM Higgs boson at the LHC. 
From Ref.~\protect\cite{gsw}.
\label{fig:Hprodcross}
}
\vspace*{-0.1in}
\end{figure}

\subsection{Higgs production channels}
\label{sec:Higgs.production}

A Higgs boson can be produced in a variety of processes at the LHC. The 
machine has sufficient energy to excite heavy quanta, and since the Higgs 
boson couples to other particles proportional to their mass, this leads to
efficient Higgs production modes. The
two dominant processes are shown in Fig.~\ref{fig:Hprodfeyn},
gluon fusion, $gg\to H$, which proceeds via a top quark loop, and weak boson 
fusion, $qq\to qqH$, where the two incoming quarks radiate two virtual $W$s 
or $Z$s which then annihilate to form the Higgs. The expected cross sections 
for both are in the 1--30~pb range, and are shown in
Fig.~\ref{fig:Hprodcross} as a function of the Higgs mass.

Beyond these two, a variety of heavy particle 
production processes may radiate a relatively light Higgs boson at an 
appreciable rate. These include $WH$ (or $ZH$) associated production,
\bq
\bar q q \to WH\;,
\eq
which is the analogue of $ZH$ production at $\epem$ colliders, and
$t\bar tH$ (or $b\bar bH$) associated production,
\bq
\bar qq \to \bar ttH\;, \qquad gg\to \bar ttH\;.
\eq
As can be seen in Fig.~\ref{fig:Hprodcross}, these associated production 
cross sections are quite small for large Higgs boson masses, but can become
interesting for $m_H\lsim 150$~GeV, because the decay products of the 
additional $W$s or top quarks provide characteristic signatures of associated
Higgs production events which allow for excellent background suppression.

In the most relevant region, 400~GeV~$\gsim m_H\gsim 110$~GeV, which will 
not be accessible by LEP, the total SM Higgs production cross section is 
of the order 10--30~pb, which corresponds to some $10^5$ events per year 
at the LHC, even at the initial 'low' luminosity of 
${\cal L}=10^{33}\;{\rm cm}^{-2}{\rm sec}^{-1}$. This already indicates that 
the main problem at the LHC is the visibility of the signal in an environment 
with very large QCD  backgrounds, and this visibility critically depends on 
the decay mode of the Higgs. The decays $H\to b\bar b,\; c\bar c,\; gg$ all 
lead to a dijet signature and are very difficult to identify, because dijet 
production at a hadron collider is such a common-place occurrence. More
promising are $H\to ZZ\to \ell^+\ell^-\ell^+\ell^-$ and $H\to\gamma\gamma$ 
which have much smaller backgrounds to contend with. 

\begin{figure}[t]
\centering\leavevmode
\psfig{figure=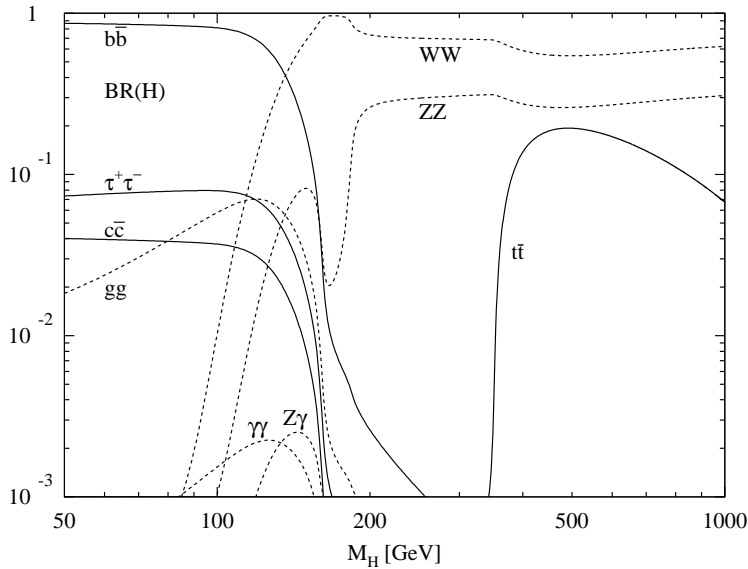,height=3.0in,angle=270}
\vspace*{-0.1in}
\caption{
Decay branching fractions of the SM Higgs boson, as a function of Higgs mass.
From Ref.~\protect\cite{spira}.
\label{fig:hdecay}
}
\vspace*{-0.1in}
\end{figure}

\subsection{Higgs search in the $H\to ZZ$ mode}
\label{sec:Higgs.to.ZZ}

The expected branching ratios of the SM Higgs to the various final states 
are shown in Fig.~\ref{fig:hdecay}. For $m_H>180$~GeV, the $H\to ZZ$ threshold,
Higgs decay to $WW$ and $ZZ$ dominates, and of the various decay modes of the
two weak bosons, $ZZ\to \ell^+\ell^-\ell^+\ell^-$ gives the cleanest 
signature.\cite{GHKD}
Not only can the invariant mass of the two lepton pairs be reconstructed, and
their arising from $Z$ decay be confirmed, also the invariant mass
of the four charged leptons reconstructs the Higgs mass. Thus, the Higgs 
signal appears as a resonance in the $m_{4\ell}$ invariant mass distribution. 
$ZZ$ backgrounds are limited at the LHC, they mostly arise from 
$\bar qq\to ZZ$ processes, i.e. continuum $ZZ$ production. The Higgs resonance 
needs to be observed on top of this irreducible background. Estimates are 
that with 100~fb$^{-1}$ the LHC detectors can see $ZZ\to 4\ell$ for 
180~GeV$<m_H\lsim 600$~GeV.\cite{ATLAS,CMS} 

The only disadvantage of this 'gold-plated' 
Higgs search mode is the relatively small branching ratio 
of $B(H\to ZZ\to \ell^+\ell^-\ell^+\ell^-)\approx 0.15\%$.
For large Higgs masses the gold-plated mode becomes rate limited, and 
additional Higgs decay modes must be searched for. 
$H\to ZZ\to\nu\bar\nu\ell^+\ell^-$, the 'silver-plated' mode, has about a 
six times larger rate, but because of the unobserved neutrinos it does not 
provide for a direct Higgs mass reconstruction. The large missing $E_T$ 
and the two observed leptons allow a measurement of the transverse mass, 
however, with a Jacobian peak at $m_H$, analogous to  the $W\to\ell\nu$ 
example. $H\to ZZ\to\nu\bar\nu\ell^+\ell^-$ events allow an extension 
of the Higgs search to $m_H\approx 0.8$--1~TeV.

Another promising search mode for a heavy Higgs boson is 
$H\to WW\to\ell\nu jj$, where the Higgs boson is produced in the weak boson
fusion process, as depicted in Fig.~\ref{fig:Hprodfeyn}(b). The two quarks
in the process $qq\to qqH$ result in two additional jets, which have a 
large mutual invariant mass, and the presence of these two jets is a very
characteristic signature of weak boson fusion processes. These two jets 
are typically emitted at forward angles, corresponding to pseudo-rapidities 
between $\pm 2$ to $\pm 4.5$. Requiring the observation of these two 
'forward tagging jets' substantially reduces the backgrounds and leads 
to an observable signal for Higgs boson masses in the 600~GeV to 1~TeV 
range and above.\cite{ATLAS,CMS,IZ} 
This technique of forward jet-tagging can more generally be used to search 
for any weak boson scattering processes.\cite{Peskin,bagger}
However, it is as useful for the study of a weakly interacting Higgs sector
at the LHC, and we shall consider it below in some detail in that context.

Fits to electroweak precision data, from LEP and SLC, from lower energy data
as well as from the Tevatron, are increasingly pointing to a relatively small
Higgs boson mass,\cite{marciano} between 100 and 200~GeV, at least within 
the context of the SM. Thus, interest at present is focused on search 
strategies for a relatively light Higgs boson. The search in the gold-plated 
channel, $H\to\ell^+\ell^-\ell^+\ell^-$, can be extended well below the 
$Z$-pair threshold.\cite{ATLAS,CMS} As can be read off Fig.~\ref{fig:hdecay}, 
the branching ratio $H\to ZZ^*$, into one real and one virtual $Z$, remains 
above the few percent level for Higgs boson masses as low as 130~GeV.
With an integrated luminosity of around 100~fb$^{-1}$ the SM Higgs resonance 
can be observed the LHC, above $m_H\approx 130$~GeV. A problematic region is
the 160~GeV$<m_H<$180~GeV range, however, where the branching ratio 
$H\to\ell^+\ell^-\ell^+\ell^-$ takes a serious dip: this is the region 
where the Higgs boson can decay into two on-shell $W$s, while the $ZZ$ 
channel is still below threshold for two on-shell $Z$s. While sufficiently 
large amounts of data will yield a positive Higgs signal in this region, the 
observation in the dominant decay channel, via
$H\to\wpwm\to \ell^+\nu\ell^-\bar\nu$ has a much higher rate 
and can in fact be distinguished from backgrounds as well.\cite{DD}

\subsection{Search in the $H\to\gamma\gamma$ channel}
\label{sec:Higgs.to.AA}

\begin{figure}[t]
\centering\leavevmode
\psfig{figure=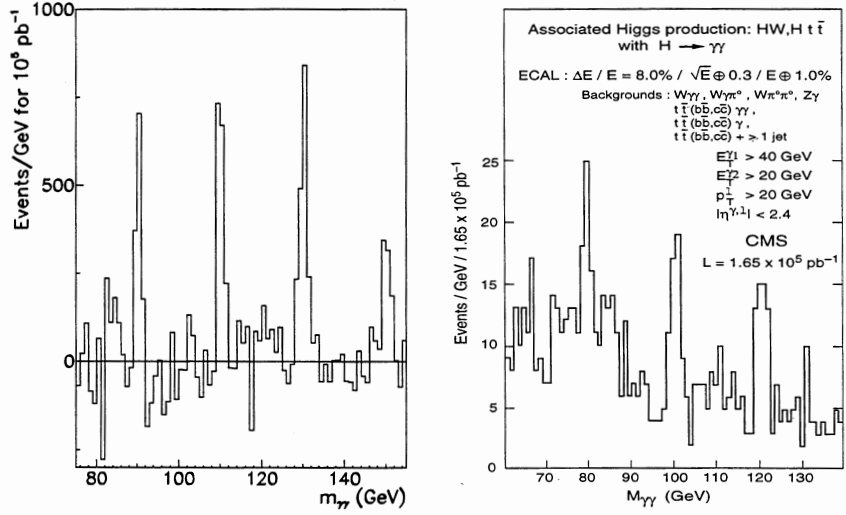,height=3.0in}
\caption{
Expected $H\to\gamma\gamma$ signal in the CMS detector at the LHC, after one 
year of running at a luminosity of $10^{34}\;{\rm cm}^{-2}{\rm sec}^{-1}$.
Shown are the expected Higgs resonance peaks in the $\gamma\gamma$ 
invariant mass distributions (a) in the inclusive $H\to \gamma\gamma$ search 
after background subtraction for Higgs masses of 90, 110, 130, and 150~GeV, 
and (b) for $WH$ and $\bar ttH$ associated production with $m_H=80$, 100,
and 120~GeV. From Ref.~\protect\cite{CMS}.
\label{fig:cmsAA}
}
\vspace*{-0.1in}
\end{figure}

For a relatively light SM Higgs boson, of mass $m_H\lsim 150$~GeV, the 
cleanest search mode is the decay $H\to\gamma\gamma$. Photon energies can be
measured with high precision at the LHC, resulting in a very good
$m_{\gamma\gamma}$ invariant mass resolution, of order 1~GeV for Higgs masses
around 100--150~GeV. In fact, the design of the LHC detectors has been 
driven by a good search capability for these events. Since the natural
Higgs width is only a few MeV in this mass range, the Higgs would appear as
as a very narrow peak in the $\gamma\gamma$ invariant mass distribution.

The simplest search is for all $H\to\gamma\gamma$ events, irrespective of the
Higgs production mode. Backgrounds arise from double photon bremsstrahlung
processes like
\bq
gq\to gq\gamma\gamma\;, \qquad qq\to qq\gamma\gamma\;, \quad {\rm etc.}
\eq
pair production of photons in $\bar qq$ annihilation and in gluon fusion 
(via quark loops),
\bq
\bar qq\to \gamma\gamma\;, \qquad\qquad  gg\to \gamma\gamma \;,
\eq
and from reducible backgrounds, in particular from jets where the bulk 
of the jet's energy is carried by a single $\pi^0$, whose decay 
into two nearby photons may not be resolved and may mimic a single photon 
in the detector. 

Bremsstrahlung photons tend to be emitted close to the parent quark
direction, i.e. they are close in angle to a nearby jet. The same is true 
for photons from $\pi^0$ decays, these photons are usually embedded into a 
hadronic jet. One therefore requires the signal photons to be well isolated,
i.e. to have little hadronic activity at small angular separations
\bq
\Delta R = \sqrt{(\eta_\gamma-\eta_j)^2 + (\phi_\gamma-\phi_j)^2 }\; .
\eq
Here $(\eta_\gamma,\;\phi_\gamma)$ give the photon direction and 
$(\eta_j,\;\phi_j)$ denotes the direction of hadronic activity, be it hadrons,
partons in a perturbative calculation, or jets. A practical requirement may be
that the total energy carried by hadrons within a cone of radius 
$\Delta R=0.3$ be less that 10\% of the photon energy and that no hard
jet is found with a separation $\Delta R<0.7$. Photon isolation
requirements drastically reduce bremsstrahlung and QCD ($\pi^0$) backgrounds 
and are absolutely crucial for identifying any signal.

Another characteristic feature of the backgrounds is their soft photon 
spectrum: background rates drop quite fast with increased photon transverse 
momentum, $p_{T\gamma}$.
This plays into the features of the signal. Since a Higgs boson decays
isotropically, with $E_\gamma=m_H/2$ in the Higgs rest frame, photon
transverse momenta in the range $m_H/4\lsim p_{T\gamma}\leq m_H/2$ are 
favored. In practice, when searching for a Higgs boson in the
100~GeV~$<m_H<$~150~GeV range, one requires $p_{T\gamma_1}>40$~GeV, 
$p_{T\gamma_2}>25$~GeV for the two photons, which together with the isolation
requirement reduces the background to a level of order 
$d\sigma/dm_{\gamma\gamma}=$100--200~fb/GeV.\cite{CMS} This needs to be
compared to the SM Higgs signal, which has a cross section, after cuts, of 
$\sigma\cdot B(H\to\gamma\gamma)\approx 40$~fb for masses around 
$m_H=120$~GeV. 

The visibility of the signal crucially depends on the mass resolution of   
of the detector. For CMS (ATLAS) one expects a resolution of order 
$\sigma_m=\pm 0.8$~GeV ($\pm 1.5$~GeV). 
Taking the better CMS resolution and an 
integrated luminosity of 50~fb$^{-1}$ as an example, one would see 
$S=0.683\cdot 2000=1400$ signal events in a mass bin of full width 1.6~GeV,
on top of a background of $B=13000\cdot 1.6=21000$ events, giving a 
statistical significance $S/\sqrt{B}$ of almost 10 standard deviations, a
very significant discovery! The expected two-photon mass spectrum, after
background subtraction, is shown in Fig.~\ref{fig:cmsAA}(a). The above example 
only conveys the rough size of the signal and background in the inclusive 
$H\to\gamma\gamma$ search. More detailed estimates for a range of Higgs
masses, including detection efficiencies, the decline of background rates
with increasing $m_{\gamma\gamma}$, and variations in the signal rate as a 
function of $m_H$ can be found in Refs.~\cite{ATLAS,CMS}.

The main disadvantage of the inclusive $H\to\gamma\gamma$ search is the 
relatively small signal size as compared to the background, $S:B\approx 1:15$
for CMS and even smaller for ATLAS. A much cleaner signal can be found
by looking for Higgs production in association with other particles, in 
particular the isolated leptons arising from $W$ decays in $WH$ and $\bar ttH$
associated production. A high $p_T$ isolated lepton is very unlikely to be
produced in most background processes with two photons, and, thus, a
signal to background ratio of about 1:1 or even better can be achieved.
The results of a simulation of the anticipated $m_{\gamma\gamma}$ spectrum 
are shown in Fig.~\ref{fig:cmsAA}(b), again for 
the CMS detector.\cite{CMS} While $S:B$ is very good, cross sections
are much lower than in the inclusive $H\to\gamma\gamma$ search and integrated 
luminosities of order 100~fb$^{-1}$ or larger are needed for a significant 
signal in these associated production channels.

\subsection{Weak boson fusion}
\label{sec:WBF}

There is a danger in relying too much on the $H\to\gamma\gamma$ decay
channel for a light Higgs boson, of course: it implicitly assumes that the 
Higgs partial decay widths are indeed as large as predicted by the SM. 
Approximately, the two-photon branching ratio is given by
\bq
B(H\to\gamma\gamma)=
{\Gamma(H\to\gamma\gamma)\over \Gamma(H\to \bar bb)+\dots}\;,
\eq
i.e. a strongly increased partial width for $H\to\bar bb$ can render the
$H\to\gamma\gamma$ channel unobservably small. This is what happens in the 
MSSM, with its two Higgs doublets, which lead to two CP-even scalars, the 
light $h$ and a heavier state, $H$. For large $\tan\beta=v_1/v_2$, the 
$b$-quark Yukawa coupling is enhanced, leading to a suppressed 
$h\to\gamma\gamma$ branching ratio over large regions of parameter 
space.\footnote{In the following, no distinction will be made between
different scalar states. $H$ generically denotes the Higgs resonance which 
is being searched for.}
One thus needs to prepare for a search in other decay channels as well. 
And even if the $H\to\gamma\gamma$ mode is observed first, the other 
channels will be needed to learn about the various couplings of the 
Higgs boson, to weak bosons, quarks and leptons, i.e. they need to be 
studied in order to understand the dynamics of the symmetry breaking sector.

In any model which treats lepton and quark mass-generation symmetrically, the 
$H\to \bar bb$ and $H\to\tau^+\tau^-$ decay widths move in unison because
both represent the isospin $-1/2$ component of a third generation doublet.
Thus, the $h\to\tau^+\tau^-$ branching ratio is fairly stable, staying at the
8--9\% level in e.g. the MSSM over large regions of parameter space where 
the $h\to \gamma\gamma$ branching ratio may be suppressed by large factors.
Interestingly, the tau decay mode is observable in the most copious of the 
associated Higgs production processes, weak boson fusion as depicted in 
Fig.~\ref{fig:Hprodfeyn}.

Traditionally, weak boson fusion has been considered mainly as a method for
studying a strongly interacting symmetry breaking sector,\cite{GHKD}
where one encounters either a very
heavy Higgs boson or non-Higgs dynamics such as in technicolor models. 
However, as is evident from Fig.~\ref{fig:Hprodcross}, the weak boson fusion
cross section, $\sigma(qq\to qqH)$, is as large as a few pb also for 
a Higgs boson in the 100~GeV range, which in the SM corresponds to 10--20\% 
of the total Higgs production rate. 

A characteristic feature of weak boson fusion events are the two 
accompanying quarks (or anti-quarks) from which the ``incoming'' $W$s or 
$Z$s have been radiated (see Fig.~\ref{fig:Hprodfeyn}(b)). In general these 
scattered quarks will give rise to hadronic jets. By tagging them, i.e. by 
requiring that they are observed in the detector, one obtains a powerful 
background rejection tool.\cite{Cahn,Froid} Whether such an 
approach can be successful depends on the properties of the tagging jets:
their typical transverse momenta, their energies, and their angular 
distributions.

Similar to the emission of virtual photons from a high energy electron 
beam, the incoming weak bosons tend to carry a small fraction of 
the incoming parton energy. At the same time the incoming 
weak bosons must carry substantial energy, of order $m_H/2$, in 
order to produce the Higgs boson. Thus the final state quarks in 
$qq\to qqH$ events typically carry very high energies, of order 
1~TeV. This is to be contrasted with their transverse momenta, 
which are of order $p_T\approx m_W$. This low scale is set by the weak boson 
propagators in Fig.~\ref{fig:Hprodfeyn}(b), which introduce a factor
\bq
D_V(q^2) = {-1\over q^2 - m_V^2} \approx {1\over p_T^2+m_V^2}
\eq
into the production amplitudes and
suppress the $qq\to qqH$ cross section for quark transverse momenta above
$m_V$. The modest transverse momentum and high energy of the scattered quark 
corresponds to a small scattering angle, typically in the $1<\eta<5$ 
pseudo-rapidity region. 

\begin{figure}[t]
\centering\leavevmode
\psfig{figure=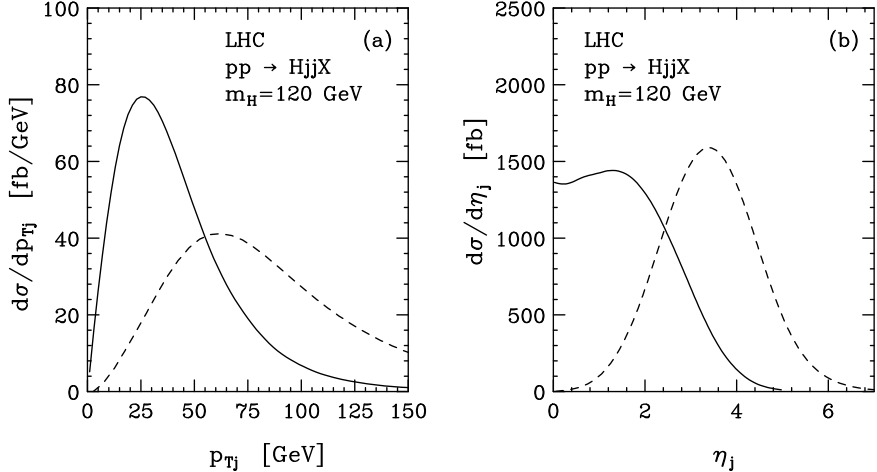,height=2.5in,angle=90}
\caption{
Transverse momentum and pseudo-rapidity distributions of the two (anti)quark 
jets in $qq\to qqH$ events at the LHC. Shown are (a) $d\sigma/dp_{Tj}$ for
the highest (dashed curve) and lowest $p_T$ jet (solid curve) and (b) 
$d\sigma/d|\eta_j|$ for the most forward (dashed curve) and the most central
jet (solid curve).
\label{fig:ptjH}
}
\vspace*{-0.1in}
\end{figure}

These general arguments are confirmed by Fig.~\ref{fig:ptjH}, where the 
transverse momentum and pseudo-rapidity distributions of the two potential 
tagging jets are shown for the production of a $m_H=120$~GeV Higgs boson at 
the LHC. One finds that one of the two quark jets has substantially lower 
median $p_T$ ($\approx 35$~GeV) than the other ($\approx 70$~GeV), and
therefore experiments must be prepared to identify fairly low $p_T$ forward 
jets. A typical requirement would be~\cite{rzh}
\bq
p_{T_{j(1,2)}} \geq 40, 20~{\rm GeV} \, ,\qquad |\eta_j| \leq 5.0 \,, \qquad
m_{j1,j2}> 1\;{\rm TeV}\;,
\eq
where, in addition, the tagging jets are required to be in opposite 
hemispheres, with the Higgs decay products between them. 

While these requirements will suppress backgrounds substantially, the most 
crucial issue is identification of the $H\to\tau\tau$ decay products and 
the measurement of the $\tptm$ invariant mass. The $\tau$'s decay inside 
the beam pipe and only their decay products, an electron or muon in the 
case of leptonic decays, $\tau^-\to\ell^-\bar\nu_\ell\nu_\tau$, and an 
extremely narrow hadronic jet for $\tau^\pm\to h^\pm\nu_\tau$ 
($h=\pi,\rho,a_1$) are seen inside the detector. The presence of a charged 
lepton, of $p_T>20$~GeV, is crucial in order to trigger on an $H\to\tptm$ 
event. Allowing the other $\tau$ to decay hadronically then yields the 
highest signal rate. Even though the hadronic $\tau$ decay is seen as a 
hadronic jet, this jet is not a typical QCD jet. Rather, the $\tau$-jet
is extremely well collimated and it normally contains a single 
charged track only (so called 1-prong $\tau$ decay). An analysis by 
ATLAS~\cite{ATLAS,Cavalli} has shown that hadronically decaying $\tau$'s, of
$p_T>40$~GeV, can be identified with an efficiency of 26\% while rejecting 
QCD jets at the 1:400 level. 

At first sight the two or more missing neutrinos in $H\to\tptm$ decays seem
to preclude a measurement of the $\tptm$ invariant mass. However, because of
the small $\tau$ mass, the decay products of the $\tau^+$ or the $\tau^-$ 
all move in the same direction, i.e. the directions of the unobserved 
neutrinos are known. Their energy can be inferred by measuring the missing 
transverse momentum vector of the event. 
Denoting by $x_{\tau_\ell}$ and $x_{\tau_h}$ the fractions of the parent 
$\tau$ carried by the observed lepton and decay hadrons, the transverse 
momentum vectors are related by
\bq
\label{eq:taurecon}
{\sla{\bf p}}_T = ({1\over x_{\tau_l}} - 1) \; {\bf p}_{T\ell} +
({1\over x_{\tau_h}} - 1) \; {\bf p}_{Th} \; .
\eq
As long as the the decay products are not back-to-back, Eq.~(\ref{eq:taurecon})
gives two conditions for $x_{\tau_i}$ and provides the $\tau$ momenta as
${\bf p}_{\ell}/x_{\tau_l}$ and ${\bf p}_h/x_{\tau_h}$, respectively.
As a result, the $\tptm$ invariant mass can be reconstructed,\cite{tautaumass}
with an accuracy of order 10--15\%.

\begin{figure}[t]
\centering\leavevmode
\psfig{figure=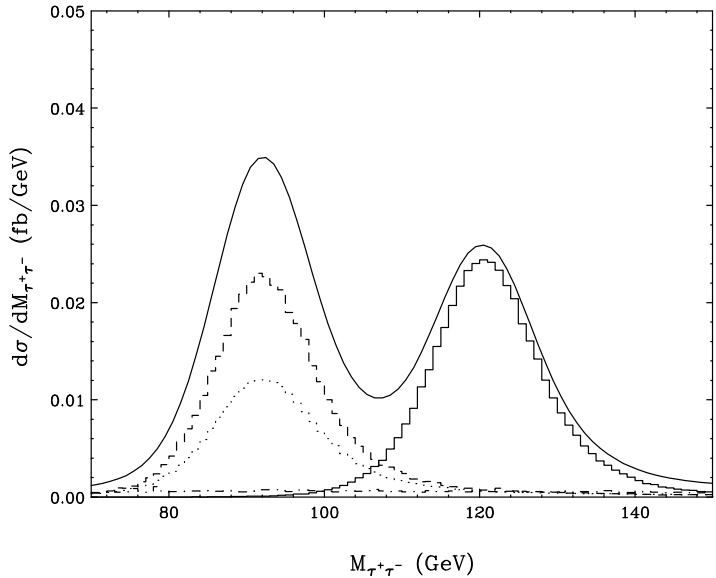,height=3.0in,angle=90}
\caption{
Reconstructed $\tptm$ invariant mass distribution expected in weak
boson fusion events at the LHC, for a SM Higgs boson of mass $m_H=120$~GeV.
The solid line represents the sum of the signal and all backgrounds.
Individual components are shown as histograms:
the $qq\to qqH$ signal (solid), the irreducible QCD $Zjj$
background (dashed), the irreducible EW $Zjj$ background (dotted), and the 
combined $Wj+jj$ and $b\bar{b}jj$ reducible backgrounds (dash-dotted).
From Ref.~\protect\cite{rzh}.
\label{fig:Mtautau}
}
\vspace*{-0.1in}
\end{figure}

Backgrounds to $H\to\tau\tau$ events in weak boson fusion arise from several
sources. First is the production of real $\tptm$ pairs in ``$Zjj$ events'', 
where the real or virtual $Z$ or photon, which decays into a $\tau$ pair,
is produced in association with two jets. 
In addition, any source of isolated leptons and three or more jets
gives a background since one of the jets may be misidentified as a $\tau$
hadronic decay. Such reducible backgrounds can arise from $W+3$~jet production
or heavy flavor production, in particular $\bar bbjj$ events, where the $W$
or one of the $b$-quarks decay into a charged lepton. 
Identifying the two forward jets of the signal, with a large separation,
$|\eta_{j_1}-\eta_{j_2}|>4.4$, and large invariant mass, substantially limits 
these backgrounds, as do the $\tau$-identification requirements. Additional
background reduction is achieved by asking for consistent values of the 
reconstructed $\tau$
momentum fractions carried by the central lepton and $\tau$-like jet,
\bq
\label{eq:x1x2}
x_{\tau_l} < 0.75 \, , \qquad   x_{\tau_h} < 1 \, .
\eq
Finally, a characteristic difference between the weak boson fusion signal
and the QCD backgrounds is in the amount and angular distribution of gluon 
radiation in the central region.\cite{bjgap} 
The $qq\to qqH$ signal proceeds without 
color exchange between the scattered quarks. Similar to photon bremsstrahlung
in Rutherford scattering, gluon radiation will be emitted in the very
forward and very backward directions, between the tagging jets and the beam
direction. Gluons giving rise to a soft jet are a rare occurrence in the
central region. The background processes, on the other hand, proceed by color
exchange between the incident partons, and here, fairly hard gluon radiation 
in the central region is quite common. A veto on any additional jet activity
between the two tagging jets, of $p_{Tj}>20$~GeV, is expected to reduce
the QCD backgrounds by about 80\% while reducing the signal by 20-30\% 
only.\cite{rzh} Combining these various techniques, forward jet tagging,
$\tau$-identification, $\tau$-pair mass reconstruction, and the central jet 
veto, one obtains a very low background signal (S:B~$\approx$~7:1 for 
$m_H=120$~GeV, significantly worse only for a Higgs which is degenerate with
the $Z$) which is large enough to give a highly significant signal with
an integrated luminosity of 30--50~fb$^{-1}$. The expected $\tau$-pair 
invariant mass distribution, for a Higgs mass of 120~GeV, is shown in 
Fig.~\ref{fig:Mtautau}.

The forward jet tagging and central jet vetoing techniques described above 
can be used for isolating any weak boson fusion signal. Another example
is the search for $H\to\gamma\gamma$ events, where the Higgs has been
produced via $qq\to qqH$. About 20-30~fb$^{-1}$ of integrated luminosity are 
sufficient to observe the SM Higgs boson in the 110--150~GeV mass range this 
way,\cite{rz} which is comparable to the inclusive 
$H\to\gamma\gamma$ search described earlier. By observing both production
channels, however, $gg\to H$ and $qq\to qqH$, separate information is obtained 
on the $H\bar tt$ and $HVV$ couplings which determine the production cross 
sections.

As should be clear from the preceding examples, a veritable arsenal of methods
is available at the LHC to search for the Higgs boson and to analyze its
properties. A single one of the methods may be sufficient for discovery of the
Higgs and measurement of its mass. However, discovery will only start a much 
more important endeavor, for which all the tools will be needed: determining 
the various couplings of the Higgs boson to gauge bosons and heavy fermions, 
answering the question whether additional particles
arise from the symmetry breaking sector, like the $H^\pm$ and pseudo-scalar
$A$ of two Higgs doublet models, and, thus, finding the dynamics which is
responsible for $SU(2)\times U(1)$ breaking in nature. New methods are still 
being developed for this purpose, and all will be needed to answer these 
fundamental questions.

\section{Conclusions}
\label{sec:conclusions}

Our discussion has touched on a number of the investigations which
can be conducted at $\epem$ and hadron colliders, or at a future muon 
collider. It is by no means complete, however. The strategies for identifying 
a supersymmetry signal at the LHC, for example, have not been discussed 
in any detail. The reader is referred to other TASI lectures for this
purpose.\cite{mssm} Some general properties of experimental possibilities
should have emerged, however. 
Both $\epem$ and hadron colliders provide the necessary energy and 
luminosity to search for new particles, and to investigate their properties, 
once they have been discovered. In these goals, the two types of machines 
complement each other. 

For the foreseeable future, hadron colliders produce the highest parton center 
of mass energies and therefore have the longest reach in producing very heavy 
objects. Their disadvantage is the large $pp$ cross section, i.e. the fact 
that any new physics signal needs to be extracted from backgrounds whose 
rates are larger by many orders of magnitude. We have studied 
several examples of how this can be done. Electroweak decays, with their 
resulting isolated photons and leptons, are crucial to identify top quarks,
heavy gauge bosons or the Higgs. But precise features of the hadronic part
of new particle production events provide equally important 
information. Examples are the multiple jets and tagged $b$-quarks in top
decays, and the two forward tagging jets in weak boson fusion events.

$\epem$ colliders benefit from much better signal to background ratios,
lack of an underlying event as encountered in hadron collisions, and the
constrained kinematics which results from the point-like character of the
beam particles: beam constraints are extremely useful tools in the 
reconstruction of events. The disadvantage of $\epem$ colliders is their
limited energy reach, of course, as compared to hadron colliders.

Only time will show which of these machines, the Tevatron, LEP2, the
LHC, an NLC, or a muon collider, will give us the most important clues to
what lies beyond the standard model. But given their capabilities, 
exciting times ahead of us are virtually assured.

\section*{Acknowledgments}
I would like to thank the organizers of TASI~98 for bringing students and
lecturers together in a most pleasant and inspiring environment.
The help of T.~Han and Y.~Pan in obtaining several figures and
other information is greatfully acknowledged. Many thanks 
go to D.~Rainwater and K.~Hagiwara for a most enjoyable collaboration which 
led to some of the results reported in Section~\ref{sec:WBF}. 
This work was supported in part by the University of Wisconsin Research
Committee with funds granted by the Wisconsin Alumni Research Foundation and
in part by the U.~S.~Department of Energy under Contract
No. DE-FG02-95ER40896.

\end{document}